\documentclass[journal,comsoc]{IEEEtran}
\usepackage[T1]{fontenc}

\ifCLASSINFOpdf
\else
   \usepackage[dvips]{graphicx}
   \graphicspath{{../eps/}}
   \DeclareGraphicsExtensions{.eps}
\fi
\usepackage{amsmath}
\usepackage[cmintegrals]{newtxmath}
\usepackage{cite}
\usepackage{amsfonts}
\usepackage{booktabs}
\usepackage{multirow}
\usepackage{subfigure}
\usepackage{tabularx}

\usepackage{algorithm}
\usepackage{algorithmic}
\usepackage{amsmath}
\usepackage{amssymb}
\usepackage{color}
\usepackage{array}
\usepackage{hyphenat}
\usepackage{url}

\begin{document}
%
\title{Service Chaining Placement Based on Satellite Mission Planning in Ground Station Networks}
%
%
%

\author{Xiangqiang~Gao,
        Rongke~Liu,~\IEEEmembership{Senior~Member,~IEEE,}
        and~Aryan~Kaushik,~\IEEEmembership{Member,~IEEE}
\thanks{X.~Gao and R.~Liu are with the School of Electronic and Information Engineering, Beihang University, Beijing 100191, China e-mail: (\{xggao, rongke\_liu\}@buaa.edu.cn).}
\thanks{A.~Kaushik is with the Department of Electronic and Electrical Engineering, University College London (UCL), London WC1E 7JE, United Kingdom e-mail: (a.kaushik@ucl.ac.uk).}}

\maketitle

\begin{abstract}
As the increase in satellite number and variety, satellite ground stations should be required to offer user services in a flexible and efficient manner. Network function virtualization (NFV) can provide a new paradigm to allocate network resources on-demand for user services over the underlying network. However, most of the existing work focuses on the virtual network function (VNF) placement and routing traffic problem for enterprise data center networks, the issue needs to further study in satellite communication scenarios. In this paper, we investigate the VNF placement and routing traffic problem in satellite ground station networks. We formulate the problem of resource allocation as an integer nonlinear programming (INLP) model and the objective is to minimize the link resource utilization and the number of servers used. Considering the information about satellite orbit fixation and mission planning, we propose location-aware resource allocation (LARA) algorithms based on Greedy and IBM CPLEX 12.10, respectively. The proposed LARA algorithm can assist in deploying VNFs and routing traffic flows by predicting the running conditions of user services. We evaluate the performance of our proposed LARA algorithm in three networks of Fat-Tree, BCube, and VL2. Simulation results show that our proposed LARA algorithm performs better than that without prediction, and can effectively decrease the average resource utilization of satellite ground station networks.
\end{abstract}


%
\IEEEpeerreviewmaketitle

\section{Introduction}
%
%
%
%
\IEEEPARstart{S}{oftware} defined network (SDN)\cite{8066287} and network function virtualization (NFV) \cite{7243304} play an important role in data center networks \cite{medhat2017service}. They can implement the separation of module functions and dedicated hardware equipments, where the module functions are referred to as virtual network functions (VNFs) and run on commodity servers \cite{kar2018energy,herrera2016resource}. Several VNFs are chained to be a service function chaining (SFC) and traffic flows in networks need to pass through the VNFs in a specific order \cite{bhamare2016a}. Within physical network resource constraints and service requirements, network service providers can flexibly place VNFs on network nodes and decide routing paths for traffic flows to optimize the operational efficiency in terms of energy consumption, resource utilization, operational cost, etc. \cite{kar2018energy,bhamare2016a,rankothge2017optimizing,bari2016orchestrating}. As new paradigms, the two technologies have a profound influence on the next generation networks \cite{8060513}.\par
\begin{figure}[tbp]
  \centering
  \includegraphics[width = \columnwidth]{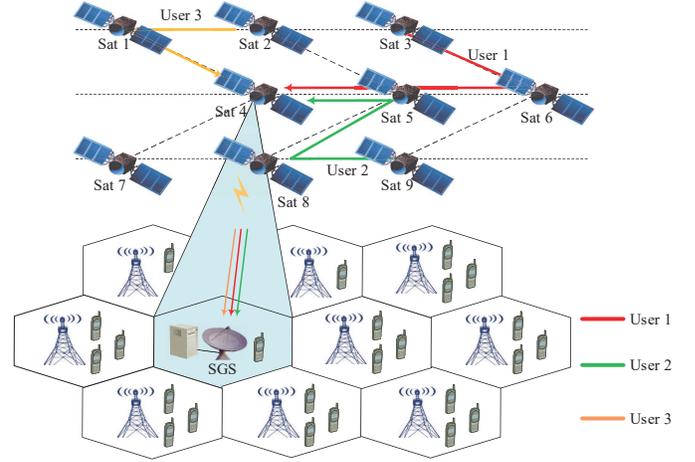}
  \caption{Procedure of running user services.}
  \label{Procedure of running user services}
\end{figure}
In satellite communication scenarios, according to the different payloads carried by satellites, satellite networks can provide various application services, e.g., remote sensing, target recognition, environmental monitoring, etc. In general, the data produced by satellite applications needs to be transmitted to a satellite ground station (SGS) network for further processing. From the perspective of SGS networks, we consider that each satellite application serves as a user and the procedure of receiving and processing the data produced by a satellite application in an SGS network serves as a user service. However, conventional SGSs consist of expensive dedicated hardware middleboxes, which are more complicated and difficult to be compatible with different user services, as the number and variety of user services increase. To provide agile service provisioning for user services, SDN and NFV are considered as new paradigms in allocating network resources on-demand and introduced to SGS networks \cite{8685771,7962772,7490612}. Similar to cloud radio access network (C-RAN), SGS networks implemented by SDN and NFV are composed of two parts: radio remote unit (RRU) and baseband processing unit (BBU), where a data center network is considered as a BBU resource pool and can deploy the VNFs concerning baseband signal processing. Multiple RRUs are connected to a BBU resource pool via high bandwidth and low latency networks. In this paper, our purpose is to focus on optimizing the resource utilization of a BBU resource pool. Thus, a user service can be viewed as a service function chaining and divided into the following function modules: (1) ingress, (2) network receiving, (3) capture, (4) tracking, (5) synchronization, (6) decoding, and (7) egress. Fig.~\ref{Procedure of running user services} describes the procedure of providing services for three users, such as user1, user2, and user3, in an SGS network. The traffic flows for the three users can be steered to an SGS network by inter-satellite links (ISLs), where three RRUs can convert the received data into baseband signal data and transmit them to a BBU resource pool for further processing, respectively.

It is worth noting that the number and type of payloads on satellites can be limited as a result of satellite's physical condition constraints \cite{7883913}. In order to effectively provide satellite application services, satellite control centers are responsible for planning satellite missions over varying times in the light of satellite's available resources and application service requirements \cite{7976296,7272824,5332322}. That is, the main difference between placing VNFs in SGS networks and other data center networks is that the user services for SGS networks are planned but for other data center networks these are uncertain. According to the results of satellite mission planning in satellite control centers, SGS networks can prior know the running periods of all satellite application services within a time frame, where the information should be used to better orchestrate VNFs to improve the performance of SGS networks. However, to the best of our knowledge, none of the existing work concerning VNF placement and resource allocation in SGS networks considered that the running periods of satellite application services can be obtained in advance, e.g., in \cite{6838778,6818281,6581256,7463884}.

In this paper, we study the problem of VNF placement and routing traffic in SGS networks. An integer nonlinear programming (INLP) model is formulated to minimize the link resource utilization and the number of servers used. To address the optimization problem, we propose location-aware resource allocation (LARA) algorithms based on Greedy \cite{liu2015improve} and IBM CPLEX 12.10 \cite{IBM-CPLEX-WEB}, respectively, according to predicting the running conditions of user services by satellite mission planning. Note that satellite mission planning for user services is out of the scope of this paper and we assume the information about service types, resource requirements, and the running periods for all user services can be known in advance. We make the experiments for three networks of Fat-Tree \cite{leiserson1985fat}, BCube \cite{guo2009bcube}, and LV2 \cite{greenberg2009vl2} with different number of servers to evaluate the performance of our proposed LARA algorithm. This paper provides the following contributions.\par
\begin{itemize}
  \item We build the problem of VNF placement and routing traffic by prior sensing the running conditions of satellite user services in SGS networks, where the information about service types, resource requirements, and the running periods of all user services could be predicted via satellite mission planning in satellite control centers.
  \item We formulate the problem of VNF placement and routing traffic as an INLP model and prove it to be NP-hard. Our aim is to minimize the resource utilization of networks.
  \item Two location-aware resource allocation algorithms based on Greedy and CPLEX are implemented to address the problem of VNF placement and routing traffic.
  \item We evaluate the performance of our proposed Greedy- and CPLEX-based LARA algorithms in BCube networks with $4$ and $8$ servers, respectively, and can observe that the proposed LARA algorithm based on CPLEX is suitable for solving the problem of resource allocation in small scale networks due to the computational complexity.
  \item Furthermore, we simulate and evaluate the performance of our proposed Greedy-based LARA algorithm for different number of predictable time slots in three networks of Fat-Tree, BCube, and LV2 with the different number of servers.
\end{itemize}

The remainder of this paper is organized as follows: Section \ref{Literature Review} briefly reviews related work about the VNF placement and routing traffic problems. Section \ref{System Model} introduces the system model of resource allocation in terms of a physical network and user services. In Section \ref{Problem Formulation}, we formulate the problem of resource allocation as an INLP model and analyze the computational complexity. Location-aware resource allocation algorithms based on Greedy and IBM CPLEX are proposed in Section \ref{Algorithm}. Section \ref{Performance Evaluation} discusses the performance of our proposed LARA algorithm in three different networks. Finally, we provide the conclusion of this paper in Section \ref{Conclusion}.\par

\section{Literature Review}\label{Literature Review}
In this section, we first discuss the VNF placement problem in general data centers. Then we introduce the related work concerning SGS network virtualization. Finally, we briefly review the existing work related to satellite mission planning.

\subsection{VNF Placement Problem}
\begin{table*}[tbp]
  \centering
  \caption{Literature Review and Comparison with the Proposed Work}
  \label{Literature Review and Comparison with the Proposed Work}
    \resizebox{\textwidth}{!}{
    \begin{tabular}{|p{0.1\textwidth}<{\centering}|p{0.29\textwidth}<{\centering}|p{0.29\textwidth}<{\centering}|p{0.22\textwidth}<{\centering}|}
    \hline
    Reference & Predictable/Unpredictable & Objective Function & Optimization Approach \\
    \hline
    \cite{rankothge2017optimizing}   & Unpredictable & Required resources & Genetic \\
    \hline
    \cite{bari2016orchestrating}   & Unpredictable & Operational cost, Resource fragmentation & CPLEX, Viterbi \\
    \hline
    \cite{liu2015improve}  & Unpredictable & End-to-end delay, Bandwidth consumption &Greedy, Simulated annealing  \\
    \hline
    \cite{raayatpanah2018virtual}  & Unpredictable & Energy consumption & CPLEX \\
    \hline
    \cite{tan2017nsga}  & Unpredictable & Service cost, Energy consumption & NSGA-II \\
    \hline
    \cite{sun2016forecast}  & Predictable, Fourier-series-based prediction & Deployment cost & Forecast-assisted online algorithm \\
    \hline
    \cite{tang2018dynamic}  & Predictable, Traffic forecasting method & Number of VMs, Cross-rack traffic & Primal-dual, Relaxation algorithm \\
    \hline
    Proposed work & Predictable, Satellite mission planning & Resource utilization & CPLEX, Greedy \\
    \hline
    \end{tabular}%
    }
  \label{tab:addlabel}%
\end{table*}%
The problem of VNF placement and routing traffic in cloud environment is demonstrated as NP-hard \cite{liu2015improve,bari2016orchestrating}. Due to the computational complexity of an ILP problem, heuristic algorithms are widely used to find an approximated solution in practical applications \cite{rankothge2017optimizing,raayatpanah2018virtual,tan2017nsga}.

The authors in \cite{rankothge2017optimizing} formulated an ILP problem to optimize the resource utilization of servers, links, and bandwidths, and used a genetic algorithm to address the resource allocation issue. In \cite{bari2016orchestrating}, the authors discussed the VNF placement problem to minimize the operational expenditure of a network and resource fragment, and proposed a viterbi algorithm to tackle the problem, where they assumed that some VNFs can only run on a particular set of servers and several SFCs can share a VNF instance. In \cite{liu2015improve}, the authors proposed two heuristic algorithms based on greedy and simulated annealing to minimize the end-to-end delay and the bandwidth consumption. The authors in \cite{raayatpanah2018virtual} presented the VNF placement problem for SFCs with minimizing energy consumption and addressed the problem with CPLEX. In \cite{tan2017nsga}, the authors considered a resource allocation problem for virtual machines and proposed a fast elitist non-dominated sorting genetic algorithm (NSGA-II) to allocate service resources in cloud.

Some of existing work discusses that the resource and workload prediction assists in improving the operational efficiency of networks \cite{sun2016forecast,tang2018dynamic,li2018deep}. A forecast-assisted SFCs placement by affiliation-aware VNF placement is presented in \cite{sun2016forecast}, where the future VNF requirements can be forecasted based on a fourier-series prediction method. In \cite{tang2018dynamic}, the authors proposed a traffic forecasting method by analyzing the traffic characteristics in data center networks and implemented two VNF placement algorithms to scale the VNF instances dynamically, where the optimization problem is formulated to minimize the number of virtual machines for deploying VNFs.

\subsection{SGS Network Virtualization}
In some of previous work  \cite{6581256,jou2018architecture,ferrus2016virtualization,ahmed2017satellite}, SDN and NFV are introduced into satellite communication to facilitate the flexibility and scalability. The authors in \cite{8685771} discussed the service function chaining placement problem in terrestrial and satellite ground clouds based on SDN and NFV for improving the resource utilization of the underlying network. In \cite{7962772}, the authors implemented an architecture of satellite ground segment systems by using SDN and NFV to address the problem of allocating satellite bandwidth resources on-demand. In \cite{7490612}, the authors discussed a virtual satellite ground station and its potential applications to reduce capital and operational expenditures. References \cite{6838778} and \cite{6818281} proposed tabu search and neighbor-area algorithms to solve the task scheduling problem of satellite ground stations, respectively. In \cite{7463884}, the authors proposed a resource mapping method based on multi-priority coefficient for providing available resources for tasks in cloud-based satellite ground systems. In \cite{ferrus2016virtualization}, the authors discussed an innovative architecture of satellite ground systems by SDN and NFV, and used the proposed architecture to dynamically orchestrate satellite communication services to improve the system flexibility and reconfigurability. In \cite{ahmed2017satellite}, the authors proposed satellite ground segment systems with SDN and NFV to implement the diversity of satellite gateway with higher capacity enhancement, failover, and resiliency management. A shared satellite ground station is proposed by using user-oriented virtualization to address complex satellite telemetry, tracking, and command (TT\&C) in \cite{9051712}.

\subsection{Satellite Mission Planning}
For earth observation satellites, satellite mission planning should be considered for improving the operational efficiency \cite{946526,7976296}. A market-based conflict resolution approach was proposed for planning earth observation missions in \cite{946526}. In \cite{8418365}, the authors formulated the agile satellite mission planning problem as a mixed integer optimization problem and addressed that by a preference-based evolutionary multiple objective optimization. In \cite{5980932}, the authors discussed an online system for planning satellite observation missions to improve the operational efficiency. The authors in \cite{8865283} discussed the problem of satellite mission planning by using a genetic algorithm.

Reviewed related work is summarized in Table \ref{Literature Review and Comparison with the Proposed Work} and comparison with our proposed work is also provided. In our paper, we investigate the VNF placement problem based on satellite mission planning in SGS networks while minimizing the resource utilization. We formulate the VNF placement problem as an INLP problem and propose location-aware resource allocation algorithms based on CPLEX and Greedy by prior obtaining the running periods of user services from satellite mission planning.

\section{System Model}\label{System Model}
In this section, we describe the system model for user services and an SGS network in detail, and discuss the problem of VNF placement for user services, where the SGS network and user services are considered as directed acyclic graphs (DAGs).\par

\subsection{User Service}
We denote the set of user services as $Q$ with $K$ user services. Each user service $q_k \in Q$, which is viewed as a service function chaining, consists of multiple VNFs in a specific order and can be expressed as a directed acyclic graph $G(F_k,H_k)$. $F_k=\{f_{k,1}=s_k,f_{k,2},\cdots,f_{k,\left |F_k \right |}=d_k\}$ denotes the VNFs in $q_k$, where $s_k$ and $d_k$ indicate the ingress and egress, respectively, and $f_{k,i}$ indicates the $i$-th VNF of $q_k$. The maximum delay time of user service $q_k$ is indicated as $t_{k,max}$ and the computing time of $f_{k,i}$ is indicated as $t_{k,i}$. $H_k$ denotes the set of edges and each edge $h_{k}^{i_1,i_2} \in H_k$ indicates that there are bandwidth demands $b_{k}^{i_1,i_2}$ between $f_{k,i_1}$ and $f_{k,i_2}$. Note that we assume that there can be various bandwidth demands for different edges. The $r$-th resource requirements of $f_{k,i}$ are denoted as $c_{k,i}^{r}$. We assume that $s_k$ and $d_k$ just route traffic flows over the underlying network, and are not required for any computing and storage resources of servers.\par

\begin{figure}[tbp]
  \centering
  \includegraphics[width = \columnwidth]{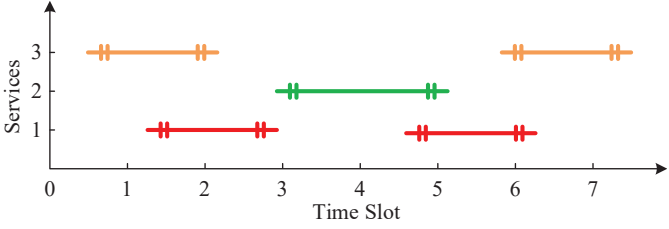}
  \caption{Life cycle time for user services.}
  \label{life cycle time for user services}
\end{figure}

In addition, each user service can be executed during a specific running period by satellite mission planning, where the start and end time for a user service is fixed. In Fig.~\ref{life cycle time for user services}, an example of the running periods for three user services is shown. It can be observed that each user service has a specific running period and can be carried out over varying times in an SGS network. We denote the running period for user service $q_k$ as $t_{k,p}$. Depending on satellite mission planning in satellite control centers, we assume that service types, resource requirements, and the running periods for user services can be prior obtained in a time frame.\par

\subsection{Physical Network}

\begin{figure}[tbp]
  \centering
  \includegraphics[width = \columnwidth]{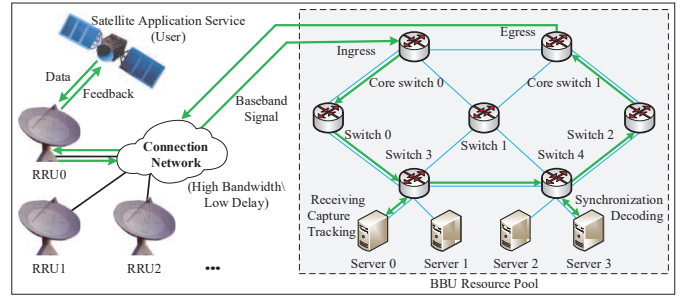}
  \caption{Example of placing VNFs in an SGS network.}
  \label{Example of placing VNFs in an SGS network}
\end{figure}

Let us denote the underlying physical network as a directed graph $G(V,E)$, where $V$ represents the set of network nodes, including servers, core switches, aggregation and edge switches, and $E$ represents the set of all links, where $L_{e}$ is the total number of links. We denote the set of servers with the number $N_{svr}$ of servers as $V_{svr}$ and the set of core switches as $V_{cs}$. The set of servers and core switches is denoted as $V_{s}=V_{svr} \cup V_{cs}$. The variable $R$ indicates the set of resources supported by servers, e.g., central processing unit (CPU), memory, and graphics processing unit (GPU). The variable $C_{n}^{r}$ is the capacity of the $r$-th resource for the $n$-th server. We assume that there are two links $(v_i,v_j)$ and $(v_j,v_i)$ between any two adjacent nodes $v_i \in V$ and $v_j \in V$. For the $l$-th link, let us denote the bandwidth capacity as $B_{l}$ and the delay time as $t_l$, respectively. Fig.~\ref{Example of placing VNFs in an SGS network} shows an example of placing VNFs for a user service in an SGS network. The SGS network consists of several RRUs and a BBU resource pool, where a connection network with high bandwidth and low delay is used between RRUs and the BBU resource pool for steering baseband signal flows to the BBU resource pool. For the BBU resource pool, there are four servers, two core switches, and five aggregation and edge switches. Different network nodes are connected with bidirectional links. A satellite transmits the data produced by a user to RRU0 in the SGS network. RRU0 converts radio front data into baseband signal data and sends them to the BBU resource pool via the connection network. In the BBU resource pool, the VNFs from the user service can be deployed on their resource requirements. The ingress and egress are on core switch0 and core switch1, respectively. The first three VNFs of network receiving, capture, and tracking are placed on server0 and the last two VNFs of synchronization and decoding are deployed on server3. The traffic flows for the user service can be described as: core switch0 $\rightarrow$ switch0 $\rightarrow$ switch3 $\rightarrow$ server0 $\rightarrow$ switch3 $\rightarrow$ switch4 $\rightarrow$ server3 $\rightarrow$ switch4 $\rightarrow$ switch2 $\rightarrow$ core switch1.\par

\subsection{SFC Placement Based on Satellite Mission Planning}

In this paper, due to satellite mission planning in satellite control centers, we assume that service types, resource requirements, and the running periods for user services in a time frame can be prior known. In order to improve the operational efficiency of an SGS network, we investigate the problem of VNF placement and routing traffic by prior sensing service types, resource requirements, and the running periods for user services.\par

For satellite communication systems, a satellite application service is performed by satellites and the produced data should be sent back to an SGS network according to satellite mission planning. The SGS network needs to provide the required resources for the user service, e.g., $q_k$, in time and deploy the VNFs on available servers to further handle the data. The ingress $s_k$ and egress $d_k$ for user service $q_k$ should be deployed on two different core switches. We place the adjacent VNFs from a user service on the same server as far as possible to save the bandwidth resources. In addition, we should further improve the resource utilization of active servers to reduce the number of servers used by user services. Our objective is to minimize the number of used servers and the link resource utilization for an SGS network. We assume that the problem of resource allocation for user services is handled in a batch processing mode. We collect the user services that are appearing in the next time slot and assign available resources to them at a specific time interval. The resource allocation algorithm is implemented based on predicting the running periods of the user services according to satellite mission planning.\par
For an SGS network, when a server is in an active state there will be the operational cost, such as energy consumption. To reduce the operational cost of an SGS network, when a server does not provide any available resources for user services and is in an idle state over a period of time, we can make the server to be in sleep or shutdown states. If the resource requirements of the current user services are more than the resource capacities of active servers, then we can wake up the servers from sleep or shutdown states to active states and provide their available resources for user services. Therefore, according to the real-time resource requirements of the current user services, we can automatically scale in or out the number of active servers for deploying the VNFs in a dynamic cloud computing environment.
\section{Problem Formulation}\label{Problem Formulation}

In this section, we provide the problem description for VNF placement and routing traffic with mathematical methods. For an SGS network, our goal is to maximize the resource utilization of active servers to save the energy cost. That is, the number of servers used by user services is as small as possible, simultaneously, we expect to minimize the resource utilization of bandwidths and links \cite{rankothge2017optimizing}. To address the problem of resource allocation, we formulate the VNF placement problem as an INLP model. The main symbols used in our problem description are summarized in Table \ref{List of Symbols}.\par

In order to better describe the problem of VNF placement and routing traffic, we denote a path between two servers or a server and a core switch as $p$. The variable $P_{n_{1},n_{2}}$ indicates the set of the $d$ shortest paths between $v_{n_1} \in V_s$ and $v_{n_2} \in V_s$. The variable $P$ is denoted as the set of all paths for each source and destination pair, which can be obtained in advance.\par

\begin{table}[tbp]
  \renewcommand{\arraystretch}{1.3}
  \caption{List of Symbols}
  \label{List of Symbols}
  \centering
  \resizebox{\columnwidth}{!}{
  \begin{tabular}{l|l}
  \hline
  \multicolumn{2}{c}{\bfseries Physical Network}\\
  \hline
  $V$ & Set of servers and all switches.\\
  $V_{svr}$ & Set of servers with the number of $N_{svr}$.\\
  $V_{cs}$ & Set of core switches.\\
  $V_{s}$ & Set of core switches and servers.\\
  $E$ & Set of $L_{e}$ links.\\
  $B_{l}$ & Bandwidth capacity of the $l$-th link.\\
  $t_{l}$ & Link delay time of the $l$-th link.\\
  $R$ & Set of resources supported by servers.\\
  $C_{n}^{r}$ & Capacity of the $r$-th resource for the $n$-th server node.\\
  $P_{n_{1},n_{2}}$ & Set of the $d$ shortest paths between $v_{n_{1}}$ and $v_{n_{2}}$.\\
  $P$ & Set of all paths from each pair of source and destination.\\
  \hline
  \multicolumn{2}{c}{\bfseries Requested Services}\\
  \hline
  $Q$ & Set of user services with the number of $K$.\\
  $q_{k}$ & The $k$-th user service.\\
  $t_{k,max}$ & Maximum delay time for user service $q_k$.\\
  $F_{k}$ & Set of virtual network functions (VNFs) offered by $q_k$.\\
  $H_{k}$ & Set of edges from $q_{k}$.\\
  $h_{k}^{i_{1},i_{2}}$ & Edge between $f_{k,i_{1}}$ and $f_{k,i_{2}}$.\\
  $f_{k,i}$ & The $i$-th VNF of the $k$-th user service.\\
  $t_{k,i}$ & Computing time for $f_{k,i}$.\\
  $s_k,d_k$ & Source and destination of the $k$-th user service.\\
  $c_{k,i}^{r}$ & The $r$-th resource requirements for $f_{k,i}$.\\
  $b_{k}^{i_{1},i_{2}}$ & Bandwidth requirements used by $h_{k}^{i_{1},i_{2}}$. \\
  \hline
  \multicolumn{2}{c}{\bfseries Binary Decision Variables}\\
  \hline
  $z_{k,i}^{n}$ & $z_{k,i}^{n} = 1$ if $f_{k,i}$ is placed on node $v_n \in V_{s}$ or $z_{k,i}^{n} = 0$.\\
  $w_{i_{1},i_{2}}^{k,p}$ & $w_{i_{1},i_{2}}^{k,p}=1$ if the path $p$ is used by $h_{k}^{i_{1},i_{2}}$ or $w_{i_{1},i_{2}}^{k,p}=0$.\\
  \hline
  \multicolumn{2}{c}{\bfseries Variables}\\
  \hline
  $x_{n}$ & $x_n=1$ if server or core switch $v_n$ is used or $x_n=0$.\\
  $y_l$ & $y_l=1$ if link $l$ is used or $y_l=0$.\\
  $e_{l}^{p}$ & $e_{l}^{p}=1$ if link $l$ is used by path $p$ or $e_{l}^{p}=0$.\\
  $U_{svr}$ & Utilization of servers.\\
  $U_{L}$ &Utilization of links.\\
  $U_{B}$ &Utilization of bandwidths.\\
  $U$ & Objective function. \\
  $\partial$ & Weight value. \\
  \hline
  \end{tabular}
  }
\end{table}
We denote a variable $x_{n}=\left \{0,1\right \}$ to represent the active state of server or core switch $v_n$.\par
\begin{equation}\label{equation1}
x_{n}=
\begin{cases}
1& \text{if server or core switch $v_{n}$ is used}, \\
0& \text{otherwise}.
\end{cases}
\end{equation}
A variable $y_{l}=\left \{0,1\right \}$ indicates whether the $l$-th link is used or not.\par
\begin{equation}\label{equation2}
y_{l}=
\begin{cases}
1& \text{if link $l$ is used}, \\
0& \text{otherwise}.
\end{cases}
\end{equation}
When two adjacent VNFs from a user service are deployed on two different servers, a path $p$ between the two servers will be selected  to route traffic flows. A variable $e_{l}^{p}$ is used to represent whether link $l$ is used by path $p$ or not.\par
\begin{equation}\label{equation3}
e_{l}^{p}=
\begin{cases}
1& \text{if link $l$ is used by path $p$}, \\
0& \text{otherwise}.
\end{cases}
\end{equation}
We define a binary decision variable $z_{k,i}^{n}=\left \{0,1\right \}$ to express whether $f_{k,i}$ is placed on server or core switch $v_n$.\par
\begin{equation}\label{equation4}
z_{k,i}^{n}=
\begin{cases}
1& \text{if $f_{k,i}$ is placed on $v_n$}, \\
0& \text{otherwise}.
\end{cases}
\end{equation}
We also define a binary decision variable $w_{i_{1},i_{2}}^{k,p}$ to indicate which path $p$ is used by edge $h_{k}^{i_{1},i_{2}}$. If path $p$ offers the traffic flows for $h_{k}^{i_{1},i_{2}}$, then  $w_{i_{1},i_{2}}^{k,p}=1$, otherwise the value is $0$.\par
\noindent For each VNF $f_{k,i} \in F_k$, it can be deployed on one and only one server or core switch $v_n \in V_s$. This constraint is represented as follows:
\begin{equation}\label{equation5}
\sum\limits_{{v_n} \in {V_{s}}} {z_{k,i}^n = 1,\forall {f_{k,i}} \in {F_k}}.
\end{equation}
In our problem formulation, we assume that the ingress and egress of each user service should be processed on two different core switches, respectively. So that we need to ensure that $s_k$ and $d_k$ for user service $q_k$ are placed on core switches. We express this constraint as follows:
\begin{equation}\label{equation6}
z_{k,i}^n \cdot (1-x_n) = 0,{f_{k,i}} = {s_k},{d_k},\forall {v_n} \in {V_{cs}}.
\end{equation}
If two adjacent VNFs from a user service are allocated on two servers or a server and a core switch, then we need to ensure that a path $p$ between the two network nodes can be provisioned. The constraint is described in equation \eqref{equation7} below.\par
\begin{equation}\label{equation7}
z_{k,{i_1}}^{{n_1}}\! \cdot \!z_{k,{i_2}}^{{n_2}} = \!\!\!\!\!\sum\limits_{p \in {P_{{n_1},{n_2}}}} \!\!\!\!\!\!{w_{{i_1},{i_2}}^{k,p}} ,\forall {v_{n_1}},{v_{n_2}} \in {V_s},{n_1} \ne {n_2},h_k^{{i_1},{i_2}} \in {H_k}.
\end{equation}
For a physical network, resource capacities of nodes and links are limited. The physical resource constraints should be guaranteed when we place VNFs to network nodes and route traffic flows between two VNFs. In this paper, we consider the resource requirements of CPU, Memory, and GPU for user services.\par
\noindent We need to ensure that the total resource requirements for user services on a physical server can not exceed its resource capacity. The resource constraint for each server is indicated as follows:
\begin{equation}\label{equation8}
\sum\limits_{{q_k} \in Q} {\sum\limits_{{f_{k,i}} \in {F_k}} {z_{k,i}^n} }  \cdot c_{k,i}^r \le {x_n} \cdot C_n^r,\forall {v_n} \in {V_{svr}},r \in R.
\end{equation}
We also need to ensure that the resource constraint for each physical link can be satisfied. The used bandwidths for a physical link should be less than its resource capacity. The related constraint for $\forall l \in E$ is depicted as follows:
\begin{equation}\label{equation9}
\left\{ {\begin{array}{*{20}{c}}
{\sum\limits_{{q_k}} {\sum\limits_{h_k^{{i_1},{i_2}}} {\sum\limits_{{v_{{n_1}}},{v_{{n_2}}}}\!{\sum\limits_p {z_{k,{i_1}}^{{n_1}}\! \cdot z_{k,{i_2}}^{{n_2}}\cdot\! w_{{i_1},{i_2}}^{k,p} \cdot e_l^p \cdot b_k^{{i_1},{i_2}}} } } } \! \le \!{y_l} \cdot {B_l}},\\
{{q_k} \in Q,h_k^{{i_1},{i_2}} \in {H_k},{v_{{n_1}}},{v_{{n_2}}} \in {V_s},{n_1} \ne {n_2},p \in {P_{{n_1},{n_2}}}}.
\end{array}} \right.
\end{equation}
When we deploy the VNFs and select paths to route traffic flows over the underlying network, the maximum delay time for a user service should be considered. We need to ensure that the source-to-destination delay time for a user service is not more than the maximum delay time. The source-to-destination delay constraint for user service $q_k$ can be expressed by:
\begin{equation}\label{equation9_1}
t_{k,execute}+ t_{k,delay} \leq t_{k,\max},
\end{equation}
where $t_{k,execute}$ is the sum of the computing time of all VNFs from $F_k$ and can be described by:
\begin{equation}\label{equation9_2}
t_{k,execute} = \sum\limits_{{f_{k,i}} \in {F_k}} {t_{k,i}},
\end{equation}
$t_{k,delay}$ is the sum of the transmission delay time of all edges from $H_k$ and can be indicated by:
\begin{equation}\label{equation9_3}
\left\{ {\begin{array}{*{20}{c}}
t_{k,delay}= \sum\limits_{h_k^{{i_1},{i_2}}} {\sum\limits_{{v_{{n_1}}},{v_{{n_2}}}} {\sum\limits_{p} {\sum\limits_{l} {z_{k,{i_1}}^{{n_1}} \cdot z_{k,{i_2}}^{{n_2}} \cdot w_{{i_1},{i_2}}^{k,p} \cdot {t_l}} } } },\\
h_k^{{i_1},{i_2}} \in {H_k}, v_{n_1},v_{n_2} \in {V_s},{n_1} \ne {n_2}, p \in P_{{n_1},{n_2}},l \in p.
\end{array}} \right.
\end{equation}

In this paper, we consider that a server can be in ON or OFF states, thus when a server is in an active state there will be an operational expenditure cost, e.g., energy consumption. We can deploy more VNFs to active servers as far as possible and improve the resource utilization of active servers. Thus, we decrease the operational expenditure cost by reducing the number of servers used by user services. In addition, when two adjacent VNFs from a user service are deployed two different network nodes, a path between the two network nodes will be used to route traffic flows through the two VNFs. To reduce the used link and bandwidth resources, we can deploy the two adjacent VNFs on the same server. For optimizing three used resources concurrently, we convert the resource optimization problem to minimizing the average resource utilization of the physical network, including servers, bandwidths, and links \cite{rankothge2017optimizing}.\par
\begin{itemize}
  \item \emph{Server utilization:} The server utilization is defined as the ratio of the number of used servers and the total number of servers in an SGS network.
  \item \emph{Link utilization:} The link utilization is defined as the ratio of the number of used links and the total number of links in an SGS network.
  \item \emph{Bandwidth utilization:} The bandwidth utilization for a link is defined as the ratio of the bandwidth resources used by user services and the total bandwidth capacity. Therefore, the bandwidth utilization in an SGS network is the average bandwidth utilization for all links.
\end{itemize}\par
\noindent The total number of active servers in the physical network is described as $\!\!\sum\limits_{{v_n} \in V_{svr}}\!\! x_n$, then the utilization $U_{svr}$ of servers can be represented as follows:
\begin{equation}\label{equation10}
U_{svr} = \frac{1}{{{N_{svr}}}} \cdot \sum\limits_{{v_n} \in {V_{svr}}} {{x_n}}.
\end{equation}
The total number of active links is expressed as $\sum\limits_{l \in E} y_l$, and the link utilization $U_L$ is indicated as follows:
\begin{equation}\label{equation11}
{U_L} = \frac{1}{{{L_e}}} \cdot \sum\limits_{l \in E} y_l.
\end{equation}
For user service $q_k$, we denote the used bandwidth resources of link $l$ as $U_{B,k}^l$. For $\forall l \in E,{q_k} \in Q$, $U_{B,k}^l$ can be expressed as:
\begin{equation}\label{equation12}
U_{B,k}^l = {\sum\limits_{h_k^{{i_1},{i_2}}} {\sum\limits_{{v_{{n_1}}},{v_{{n_2}}}}{\sum\limits_p {z_{k,{i_1}}^{{n_1}} \cdot z_{k,{i_2}}^{{n_2}}\cdot w_{{i_1},{i_2}}^{k,p} \cdot e_l^p \cdot b_k^{{i_1},{i_2}}} } }},
\end{equation}
where $h_k^{{i_1},{i_2}} \in {H_k},{v_{{n_1}}},{v_{{n_2}}} \in {V_s},{n_1} \ne {n_2},p \in {P_{{n_1},{n_2}}}$. Then the total bandwidth utilization $U_{B,Q}^l$ for link $l$ can be described as follows:
\begin{equation}\label{equation13}
U_{B,Q}^l = \frac{1}{{{B_l}}} \cdot \sum\limits_{{q_k} \in Q} {U_{B,k}^l} ,\forall l \in E.
\end{equation}
\noindent Based on the above discussion, the total bandwidth utilization $U_B$ in the physical network is represented as follows:
\begin{equation}\label{equation14}
{U_B} = \frac{1}{{{L_e}}} \cdot \sum\limits_{l \in E} {U_{B,Q}^l}.
\end{equation}
Our objective function $U$ can be expressed as a weighted sum of $U_{svr}$, $U_L$, and $U_B$ \cite{rankothge2017optimizing}.
\begin{equation}\label{equation15}
U = {\partial _{svr}} \cdot {U_{svr}} + {\partial _L} \cdot {U_L} + {\partial _B} \cdot {U_B},
\end{equation}
\noindent where $\partial _{svr}$, $\partial _L$, and $\partial _B$ are the weight factors, which can be used to adjust the preferences of different resources. We consider that ${\partial _{svr}} + {\partial _L} + {\partial _B} = 1$.
The problem of VNF placement and routing traffic is formulated as an INLP problem and the objective is to minimize the resource utilization of the underlying network with the physical resource constraints. It can be described as follows:
\begin{equation}\label{equation16}
\begin{aligned}
\text{min}\quad & U \\
s.t.\quad & \eqref{equation1}-\eqref{equation9_3}.
\end{aligned}
\end{equation}
In the next subsection, we discuss the complexity analysis of the resource allocation problem.
\subsection{Complexity Analysis}

The problem of resource allocation in equation \eqref{equation16} can be seen as NP-hard due to the fact that a single source capacitated facility location problem (SSCFLP) \cite{ahuja2004multi} can be reduced to our formulated problem.\par

For SSCFLP, there are pre-specified sites $J$ and customers $I$, respectively. The operational cost is denoted as $f_i$ and the transportation cost for customer $j$ is denoted as $c_{i,j}$ when a facility is located at a site $i$. The capacity of a facility at a site $i$ is defined by $s_i$, and the demand of customer $j$ is defined by $w_j$. A binary variable $y_i$ indicates whether a facility is located at site $i$. A binary variable $x_{i,j}$ represents whether the demand of customer $j$ is offered by a facility at site $i$. The problem of SSCFLP can be described as follows \cite{ahuja2004multi}:
\begin{equation}\label{equation17}
\begin{aligned}
\text{min}\quad & \sum\limits_{i \in I} {\sum\limits_{j \in J} {{c_{ij}} \cdot {x_{ij}}}  + \sum\limits_{i \in I} {{f_i} \cdot {y_i}} } \\
s.t.\quad & \sum\limits_{i \in I} {{x_{ij}}}  = 1,\forall j \in J,\\
\quad & \sum\limits_{j \in J} {{w_j} \cdot {x_{ij}}}  \le {s_i} \cdot {y_i},\forall i \in I,\\
\quad & {x_{ij}} \in \{ 0,1\} ,{y_i} \in \{ 0,1\} ,\forall i \in I,j \in J.
\end{aligned}
\end{equation}

In order to reduce SSCFLP to the problem of VNF placement and routing traffic in this paper, we need to redescribe our optimization problem of resource allocation. Similar to reference \cite{bari2016orchestrating}, a user service is represented as $facility\rightarrow customer$, where all VNFs from user service $q_k$ except $d_k$ are regarded as a commodity to run in a facility and $d_k$ is a customer. We set a server to be a facility and the resource capacity of a server is equal to the capacity of a facility. The resource demand of a user service on a server can be described as the demand of a customer in a facility. In addition, the resource utilization of a server represents the running cost for a facility. The used links and bandwidths for a user service can be indicated as the transportation cost from a facility to a customer. Further, we make a customer for user service $q_k$ locate on a core switch that is used by $d_k$, and path $p$ is used to route traffic flows. We ensure that the used bandwidth resources for each link are not limited. Then we can transform SSCFLP to the problem of VNF placement and routing traffic. SSCFLP is well-known as NP-hard, so the problem of resource allocation in this paper is also NP-hard.\par

\section{Proposed Algorithms}\label{Algorithm}
\begin{figure}[tbp]
  \centering
  \includegraphics[width = 0.9\columnwidth]{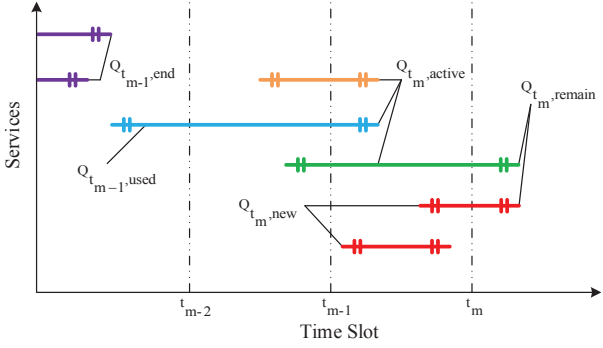}
  \caption{Procedure for running LARA algorithm.}
  \label{Procedure for running LARA algorithm}
\end{figure}
As the problem of resource allocation is NP-hard, to optimize the resource utilization, we propose two location-aware resource allocation algorithms based on Greedy and CPLEX, respectively. Firstly, we implement the location-aware resource allocation algorithm by IBM CPLEX solver with version 12.10. However, with the increase in the number of user services and scale of network, the computational complexity of solving the NP-hard problem by CPLEX increases rapidly and we must take a long computational time for addressing the problem of resource allocation. So the proposed LARA algorithm based on CPLEX is not suitable to be used in large scale problems of resource allocation. In order to solve the VNF placement and routing traffic in large scale problems, we also achieve the location-aware resource allocation algorithm based on Greedy to obtain an approximate solution.\par

\subsection{Location-Aware Resource Allocation Algorithm}\label{Location-Aware Resource Allocation Algorithm}
For an SGS network, we can prior know the information about service types, resource requirements, and life cycle time for user services in a time frame depending on satellite mission planning in satellite control centers. In view of predictable user services, we propose the location-aware resource allocation algorithm to effectively reduce the resource utilization of an SGS network in terms of servers, bandwidths, and links.\par

The procedure of resource allocation in a time slot is divided into two parts as: (1) finding an optimization solution and (2) VNF placement and routing traffic. At the beginning of a time slot, the proposed LARA algorithm is used to seek an optimization solution of resource allocation. As the results of the optimization solution, we can deploy the VNFs and select paths to route traffic flows for the current user services. The total time of the two procedures should be less than a time slot interval. For our proposed LARA algorithm, when we look for an optimization solution of resource allocation, we can predict the resource requirement and running state information about user services in the future multiple time slots according to satellite mission planning. Our purpose of resource allocation is to minimize the resource utilization in the predictable time slots as far as possible.  \par

\begin{algorithm}[tbp]
  \caption{Location-Aware Resource Allocation Algorithm.}
  \label{Location-Aware Resource Allocation}
  \hspace*{0.02in} {\bf Input:} Time slot $t$, number of predictable time slots $M$;\\
  \hspace*{0.02in} {\bf Output:} Feasible solution;
  \begin{algorithmic}[1]
  \STATE \textbf{Initialize:} $m=M$,$Q_{t_m,remain}=null$;
  \WHILE {$m>0$}
  \STATE $t_m \leftarrow t+m$;
  \STATE Obtain new user services $Q_{t_m,new}$ to be allocated resources in time slot $t_m$;
  \STATE Find all active services $Q_{t_m,active}$ at the beginning of time slot $t_m$;
  \STATE Get active services $Q_{t_{m-1},used}$ that are offered resources before time slot $t_{m}$;
  \STATE Acquire services $Q_{t_{m-1},end}$ that are finished before time slot $t_m$;
  \STATE $Q_{t_m-1,remain} \leftarrow Q_{t_m,active}-Q_{t_{m-1},used}$;
  \STATE $Q_{t_m,allocate}\!\! \leftarrow \!\!\left\{Q_{t_m-1,remain},Q_{t_m,new}\right\} - Q_{t_m,remain}$;
  \STATE Free server and bandwidth resources used by $Q_{t_{m-1},end}$;
  \STATE Allocate the resources of servers and links for $Q_{t_m,remain}$;
  \STATE Search an optimization solution of resource allocation for $Q_{t_m,allocate}$ by $Greedy$ or $CPLEX$;
  \STATE $m \leftarrow m-1$;
  \STATE $Q_{t_m,remain} \!\!\leftarrow \!\!\left\{Q_{t_m-1,remain},Q_{t_m,remain}\right\}-Q_{t_m,new}$;
  \ENDWHILE
  \RETURN Optimization solution for $Q_{t,new}$;
  \end{algorithmic}
\end{algorithm}

Fig.~\ref{Procedure for running LARA algorithm} shows the procedure for running our proposed LARA algorithm in predictable time slot $t_m$. We denote the current time slot as $t$ and the predictable time slot as $t_m$. All active user services are classified into five types according to their running states in different time slots and described as follows:
\begin{itemize}
  \item \emph{Service-type1:} For predictable time slot $t_m$, if user services in active states are over before time slot $t_m$ and there is no potential resource conflict between these user services and new user services to be allocated, we can indicate them by $Q_{t_{m-1},end}$ and the user services from $Q_{t_{m-1},end}$ are considered as service-type1.

  \item \emph{Service-type2:} At the beginning of time slot $t_m$, the user services that are still active are considered as service-type2 and denoted by $Q_{t_m,active}$.

  \item \emph{Service-type3:} The user services that are assigned network resources before time slot $t_{m}$ are considered as service-type3 and represented by $Q_{t_{m-1},used}$.

  \item \emph{Service-type4:} $Q_{t_m,new}$ indicates the user services that are occuring in time slot $t_m$. Let us denote user services in $Q_{t_m,new}$ as service-type4.

  \item \emph{Service-type5:} $Q_{t_m,remain}$ expresses the user services that are allocated network resources during $[t,t_m]$ time slots and also active in time slot $t_{m+1}$. Let $Q_{t_m,remain}$ be service-type5.

\end{itemize}\par
\begin{algorithm}[tbp]
  \caption{Greedy Algorithm.}
  \label{Greedy algorithm}
  \hspace*{0.02in} {\bf Input:} User services $Q_{t_m,allocate}$;\\
  \hspace*{0.02in} {\bf Output:} Feasible solution;
  \begin{algorithmic}[1]
  \STATE Collect active servers $V_{svr,active}$ and idle servers $V_{svr,idle}$;
  \FOR {each $q_k \in Q_{t_m,allocate}$}
  \STATE $flag, server \leftarrow Search(q_k,V_{svr,active})$;
  \IF {$flag =false$}
  \STATE $flag, server \leftarrow Search(q_k,V_{svr,idle})$;
  \STATE Add $server$ to $V_{svr,active}$;
  \STATE Remove $server$ from $V_{svr,idle}$;
  \ENDIF
  \ENDFOR
  \RETURN Feasible solution for $Q_{t_m,allocate}$;
  \end{algorithmic}
\end{algorithm}
\noindent Based on the above discussion, $Q_{t_m-1,remain}$ can be obtained by:
\begin{equation}\label{equation18}
Q_{t_m-1,remain} = Q_{t_m,active}-Q_{t_m,used},
\end{equation}
\noindent then we can obtain the user services $Q_{t_m,allocate}$ that need to be assigned in time slot $t_m$ as follows:
\begin{equation}\label{equation19}
Q_{t_m,allocate}=\left\{Q_{t_m-1,remain},Q_{t_m,new}\right\} \!-\! Q_{t_m,remain}.
\end{equation}
\noindent To effectively improve the resource utilization, we free the network resources used by user services in $Q_{t_m,end}$ and deploy the available network resources to the user services in $Q_{t_m,remain}$ by the results of resource allocation that were computed in time slot $t_{m+1}$. Then the Greedy and CPLEX approaches are carried out to find an optimization solution of resource allocation for the user services in $Q_{t_m,allocate}$. After that, $Q_{t_m,remain}$ can be updated by:
\begin{equation}\label{equation20}
Q_{t_m,remain} \!=\!\left\{Q_{t_m-1,remain},Q_{t_m,remain}\right\}\!-\!Q_{t_m,new}.
\end{equation}
The procedure of our proposed LARA algorithm is described in Algorithm \ref{Location-Aware Resource Allocation}. The current time slot is $t$ and the number of predicted time slots is $M$. At the beginning, we set $m=M$ and $Q_{t_m,remain}=null$. For time slot $t_m$, we can firstly predict $Q_{t_{m-1},end}$, $Q_{t_m,remain}$, and $Q_{t_m,allocate}$, respectively. Then we free the network resources used by user services in $Q_{t_{m-1},end}$, and allocate resources to user servers in $Q_{t_m,remain}$. Greedy and CPLEX algorithms are executed to find an optimization solution of resource allocation for user services in $Q_{t_m,allocate}$. The procedure of our proposed LARA algorithm can be executed $M$ times and then we can obtain an optimization solution of resource allocation for $Q_{t,new}$.\par

For the proposed LARA algorithm based on CPLEX, we address the INLP problem of resource allocation by IBM CPLEX solver with version 12.10, which is configured by default algorithm parameters and can obtain a global optimization solution of resource allocation.\par
In the following subsection, we discuss the Greedy algorithm used by our proposed LARA algorithm.\par
\subsection{Greedy Algorithm}\label{Greedy Algorithm}

\begin{algorithm}[tbp]
  \caption{Search.}
  \label{Search}
  \hspace*{0.02in} {\bf Input:} User service $q_k$, collection of servers $\tilde{V}_{svr}$;\\
  \hspace*{0.02in} {\bf Output:} $success, server$;
  \begin{algorithmic}[1]
  \STATE $success=false,server=null$;
  \FOR {each $v_n \in \tilde{V}_{svr}$}
  \STATE Obtain the VNF sequence $\Gamma_{k}$ of $q_k$ using a topological sort method;
  \FOR {each $f_{k,i} \in \Gamma_{k}$}
  \IF{$f_{k,i} \notin [s_k,d_k]$}
  \STATE Attempt to place $f_{k,i}$ to server $v_n$;
  \IF {$v_n$ can not offer available resources for $f_{k,i}$}
  \STATE Break;
  \ENDIF
  \ELSE
  \STATE $v_n$ is updated as a core switch used by $s_k$ or $d_k$;
  \ENDIF
  \STATE Get all predecessors of $f_{k,i}$ and their edges $H_{k,i}^{pre}$;
  \FOR {each $h_{k}^{\tilde{i},i} \in H_{k,i}^{pre}$}
  \STATE Find server $v_{\tilde{n}}$ used by $f_{k,\tilde{i}}$;
  \STATE Sort $p_{\tilde{n},n}$ between $v_{\tilde{n}}$ and $v_n$ by the path distance;
  \FOR {each $p \in p_{\tilde{n},n}$}
  \STATE Calculate available bandwidths for $h_{k}^{\tilde{i},i}$;
  \IF {there are enough bandwidths for $h_{k}^{\tilde{i},i}$}
  \STATE Break;
  \ENDIF
  \ENDFOR
  \ENDFOR
  \ENDFOR
  \IF {$q_k$ can be deployed to $v_n$}
  \STATE Perform objective function $U$;
  \IF {the objective value is better than others}
  \STATE $server=v_n$;
  \ENDIF
  \STATE $success=true$;
  \ENDIF
  \ENDFOR
  \RETURN $success, server$;
  \end{algorithmic}
\end{algorithm}
In this paper, our proposed LARA algorithm is implemented by Greedy to address the problem of resource allocation. The processing of Greedy algorithm is shown in Algorithm \ref{Greedy algorithm}. The input parameters are user services $Q_{t_m,allocate}$. At the beginning, we divide all available servers in the physical network into two portions. One is that the servers used by user services are indicated as $V_{svr,active}$. the other is that the servers in idle states are indicated as $V_{svr,idle}$. For user service $q_k \in Q_{t_m,allocate}$, we firstly call function $Search$, which will be discussed in detail later, to seek a feasible solution from servers in $V_{svr,active}$ to minimize the resource utilization. If any server in $V_{svr,active}$ can not be used by $q_k$, then $flag = false$, otherwise $flag = true$. When $flag = false$ we will find a feasible solution from servers in $V_{svr,idle}$ by function $Search$. If a server in $V_{svr,idle}$ is selected to deploy user service $q_k$, the server should be moved from $V_{svr,idle}$ to $V_{svr,active}$ and it will be in an active state. When all user services in $Q_{t_m,allocate}$ are assigned to the physical network, the Greedy algorithm will return a feasible solution. Note that we assume that an SGS network can provide enough available resources for all user services.\par

Function $Search$ is designed to deploy the VNFs on servers, and select paths to route traffic flows for the edge between two adjacent VNFs on different nodes. The aim is to minimize the resource utilization of servers, links, and bandwidths. The input parameters include user service $q_k$ and a set $\tilde{V}_{svr}$ of servers. The output parameters are an identification ``$success$'' of success and a server ``$server$'' used by $q_k$.\par

Initially, we set $success=false$ and $server=null$. For each server $v_n \in \tilde{V}_{svr}$, we attempt to deploy $q_k$ to server $v_n$. Firstly, the sequence $\Gamma_{k}$ of VNFs for $q_k$ is obtained by a topology sort method to ensure that source $f_{k,i_1}$ comes before sink $f_{k,i_2}$ for edge $(f_{k,i_1},f_{k,i_2})$. For each VNF $f_{k,i} \in \Gamma_{k}$, we place VNF $f_{k,i}$ to server $v_n$. If server $v_n$ can not satisfy the resource demands of $f_{k,i}$, then we will break the loop and turn to the next server to deploy $q_k$, otherwise we will obtain all predecessors of $f_{k,i}$ and the edges $H_{k,i}^{pre}$ between $f_{k,i}$ and its predecessors. For each edge $h_{k}^{\tilde{i},i} \in H_{k,i}^{pre}$, we search the host server $v_{\tilde{n}}$ for $f_{k,\tilde{i}}$, and sort all paths in $p_{\tilde{n},n}$ by the path distance. Then we calculate available bandwidths of each path $p \in p_{\tilde{n},n}$ for edge $h_{k}^{\tilde{i},i}$. If the bandwidth demands of edge $h_{k}^{\tilde{i},i}$ are not offered by any path $p \in p_{\tilde{n},n}$, the loop is also broken. When service $q_k$ can be deployed to server $v_n$, the objective function will be performed. If the objective value for server $v_n$ is smaller than that of others, then $server=v_n$ and $success=true$. Function $Search$ is described in Algorithm \ref{Search}.\par

\subsection{LARA Algorithm in a Dynamic Environment}\label{LARA Algorithm in Dynamic Environment}
In this paper, we allocate the available resources of an SGS network to user services on-demand by the proposed LARA algorithm in a dynamic cloud computing environment. A batch processing mode is applied to deploy user services to an SGS network. For each time slot, there are several new user services to start and some old user services to end. According to satellite mission planning, we can prior know the running periods of user services. Thus, the information concerning the new user services to appear and the old user services to end in each time slot can be obtained by an SGS network in advance. At the beginning of the current time slot, we collect the new user services that are appearing in the next time slot and after a fixed time interval, perform the proposed LARA algorithm to obtain an approximate solution of allocating the resources of an SGS network to user services. The resources used by the old completed user services can be freed into the resource pool to be available and then we can provide the available resources from the resource pool for the new user services based on that approximated solution. If the available resources of the resource pool do not fulfill the resource requirements of the current user services, the servers in sleep or shutdown states will be active to provide their available resources for user services. Moreover, when the servers are in idle states over a period of time, we can convert their states into sleep or shutdown states to reduce the operational cost of an SGS network.

\section{Performance Evaluation}\label{Performance Evaluation}

In this section, we make the experiments to evaluate the performance of the proposed LARA algorithms based on Greedy and IBM CPLEX 12.10, respectively. In small scale networks, we discuss the solution quality and computational complexity of our proposed Greedy- and CPLEX-based LARA algorithms in addressing the problem of VNF placement and routing traffic. Furthermore, we evaluate the performance of the proposed Greedy-based LARA algorithm for different predictable time slots in large scale networks. The experimental platform is a commodity server, which includes i7-4790K CPU, 16 GB of Memory, and Windows 10. PYTHON is used as our programming language.\par

\subsection{Simulation Setup}\label{Simulation Setup}
\begin{table*}[tbp]
  \centering
  \caption{Parameter Settings for Performance Evaluation}
  \label{Parameter Settings for Performance Evaluation}
    \resizebox{\textwidth}{!}{
    \begin{tabular}{|p{0.12\textwidth}<{\centering}|p{0.12\textwidth}<{\centering}|p{0.12\textwidth}<{\centering}|p{0.12\textwidth}<{\centering}|p{0.12\textwidth}<{\centering}|p{0.12\textwidth}<{\centering}
    |p{0.12\textwidth}<{\centering}|}
    \hline
    \multicolumn{7}{|c|}{Network architectures} \\
    \hline
    Topology & \multicolumn{2}{c|}{Fat-Tree} & \multicolumn{2}{c|}{BCube} & \multicolumn{2}{c|}{VL2} \\
    \hline
    Number of servers & \multicolumn{2}{c|}{16,32,48,64} & \multicolumn{2}{c|}{4,8,16} & \multicolumn{2}{c|}{16} \\
    \hline
    \multicolumn{7}{|c|}{Resource capacities for servers} \\
    \hline
    Name  & \multicolumn{2}{c|}{vCPU} & \multicolumn{2}{c|}{Memory} & \multicolumn{2}{c|}{GPU} \\
    \hline
    Capacity & \multicolumn{2}{c|}{96} & \multicolumn{2}{c|}{112 GB} & \multicolumn{2}{c|}{12} \\
    \hline
    \multicolumn{7}{|c|}{Resource capacities for links} \\
    \hline
    Name  & \multicolumn{3}{c|}{Link between a server and a switch} & \multicolumn{2}{c|}{Link between switches} & Link delay \\
    \hline
    Capacity & \multicolumn{3}{c|}{1 Gbps} & \multicolumn{2}{c|}{10 Gbps} & 0.05 ms \\
    \hline
    \multicolumn{7}{|c|}{Configurations for user services} \\
    \hline
    Name  & vCPU  & Memory & GPU   & Throughput & Delay time & Maximum delay \\
    \hline
    Network receiving & 6     & 9 GB  & 0     & 100 Mbps & 20 ms & \multirow{5}{*}{$\leq$ 1.8 s} \\
\cline{1-6}    Capture & 7     & 11 GB & 1     & 100 Mbps & 1.5 s &  \\
\cline{1-6}    Tracking & 9     & 12 GB & 1    & 100 Mbps & 100 ms &  \\
\cline{1-6}    Synchronization & 14    & 12 GB & 1    & 100 Mbps & 10 ms &  \\
\cline{1-6}    Decoding & 3     & 5 GB  & 1     & 100 Mbps & 25 ms &  \\
    \hline
    \end{tabular}%
    }
\end{table*}%

In our performance evaluation, the weight values in equation \eqref{equation15} are set as $\partial _{svr}=\partial _L=\partial _B =\frac{1}{3}$. The time slot interval is $10$ minutes. Similar to reference \cite{rankothge2017optimizing}, three network structures of Fat-Tree, BCube, and VL2 are considered to run our experiments. The main parameter settings used in the performance evaluation are listed in Table \ref{Parameter Settings for Performance Evaluation}.\par
\begin{itemize}
  \item \emph{Fat-Tree:} Fat-Tree \cite{leiserson1985fat} is a layered-structure network with core layer, aggregation layer and top-of-rack layer, and can be widely used in data center networks. A $k$ fat-tree network indicates that there are $k$ ports for each switch. It consists of $(\frac{k}{2})^{2}$ core switches and $k$ pods, where each pod include $k$ switches \cite{rankothge2017optimizing}.

  \item \emph{BCube:} BCube is a server-centric network structure for shipping-container based modular data centers. Each server has several switch ports and can connect to multiple switches of different levels. For ${BCube}_0$, $n$ servers connect to a switch with $n$ ports. A ${BCube}_k(k \geq 1)$ is constructed by $n$ ${BCube}_{k-1}s$ and $n^k$ switches with $n$ ports. There are $n^{k+1}$ servers and $k+1$ levels of switches for ${BCube}_{k}$ \cite{guo2009bcube}.

  \item \emph{VL2:} VL2 is a scalable and flexible network to support large data centers that are uniform high capacity between servers and can achieve performance isolation between services. It is composed of server layer and switch layer. Servers are connected to the switch layer by top-of-rack switches. A complete bipartite graph is formed by the links between aggregation and intermediate switches \cite{greenberg2009vl2}. For $k$-port aggregation switches and $n$-port top-of-rack switches, VL2 consists of $n\cdot \frac{k^2}{4}$ servers.

\end{itemize}\par
Due to the computational complexity of solving an INLP problem by CPLEX, the effectiveness of our proposed LARA algorithms based on Greedy and CPLEX is demonstrated in small BCube networks with 4 and 8 servers. Then we evaluate the performance of the proposed Greedy-based LARA algorithm for various predictable time slots in three networks of Fat-Tree, BCube, and VL2, where the number of servers is $16$. Furthermore, we make the experiments for Fat-Tree networks with $32,48$, and $64$ servers to evaluate the performance of our proposed Greedy-based LARA algorithm as the number of servers increases. We assume that all servers have the same resource configurations. The resource capacities for each server are $96$ vCPUs, $112$ GB Memory, and $12$ GPUs. We set the bandwidth capacity for each link between a server and a switch as $1$ Gbps and for each link between switches as $10$ Gbps \cite{greenberg2009vl2}. The link delay for each link is $0.05$ ms. The $d=8$ shortest paths between a core switch and a server or two servers can be computed in advance.\par

To simplify our simulation experiments, we assume that all satellite application services aim at observing fixed objectives on the ground by Low Earth Orbit (LEO) earth observation satellites and the fixed objectives are randomly generated. We denote the number of observation objectives by $K_{obj}$. Due to the regular orbital periods of satellites, when a satellite passes over an objective, then the objective can be observed once and the data produced by the satellite application service will be transmitted to an SGS network through inter-satellite links for further processing. Therefore, we can define a user service as the procedure of receiving and processing the downloaded data produced by a satellite application service in an SGS network, where the satellite application service is to observe a fixed objective on the ground once. The running periods for all user services can be obtained by the Satellite Tool Kit (STK) and prior known for an SGS network. According to reference \cite{2017arXiv170200369R}, we build the realistic dataset for evaluating the performance of SGS networks based on the satellite communication scenarios and reasonable estimations. We assume that each user service includes five VNFs, i.e., network receiving, capture, tracking, synchronization, and decoding, except the source and the destination. The computing time and required resources of vCPU, Memory, and GPU are different for the five VNFs, where the resource and service requirements of the VNFs are shown in Table \ref{Parameter Settings for Performance Evaluation}. In addition, we assume that the bandwidth requirements of all VNFs for each user service are the same and the value of bandwidth is $100$ Mbps. The maximum delay time for each user service is set as $1.8$ seconds.\par
\begin{figure*}[tbp]
  \centering
  \subfigure[LARA algorithm by CPLEX]{\includegraphics[width=0.28\textwidth]{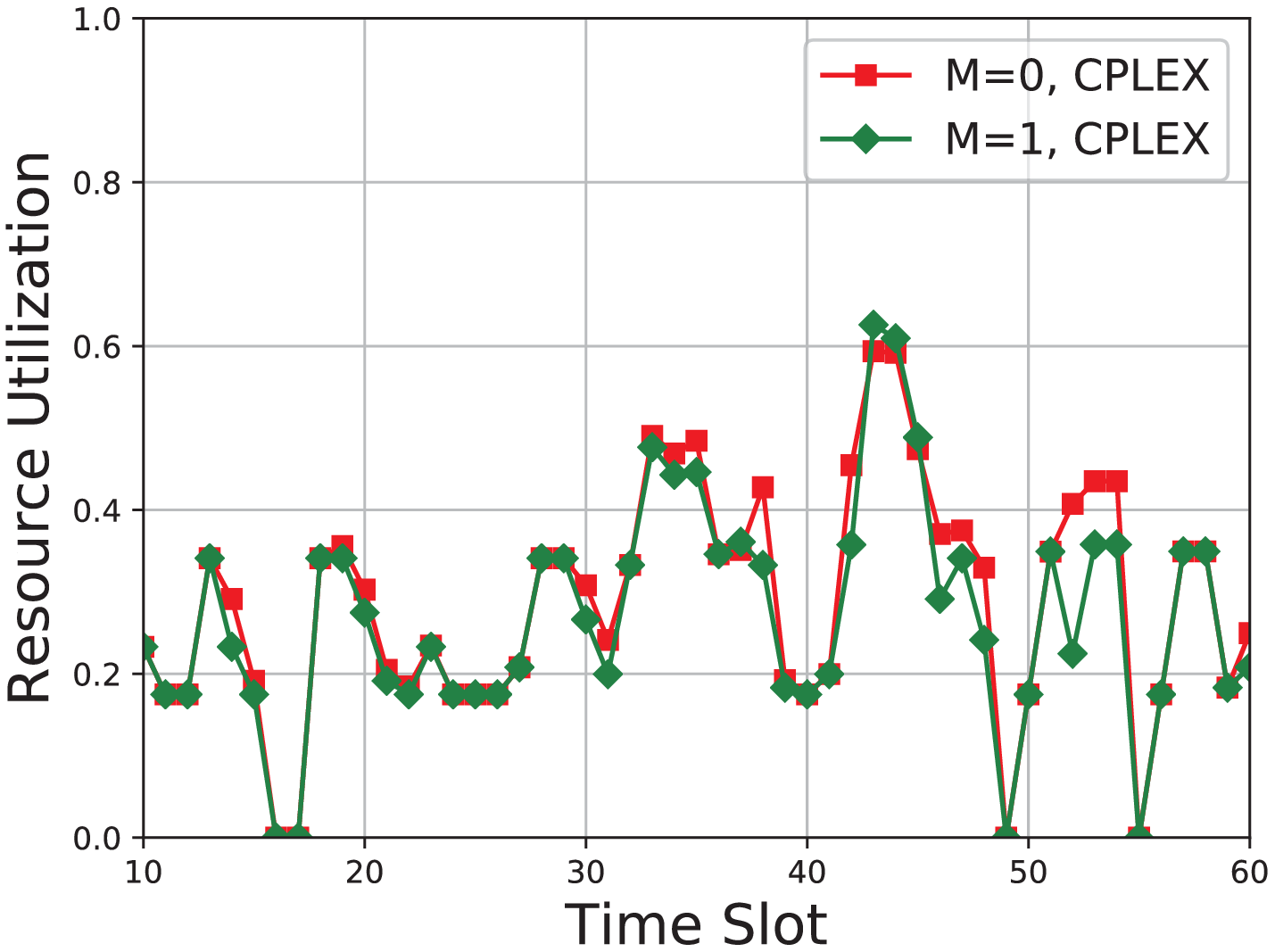}
  \label{LARA algorithm by CPLEX in BCube network with $4$ servers}}
  \subfigure[LARA algorithm by Greedy]{\includegraphics[width=0.28\textwidth]{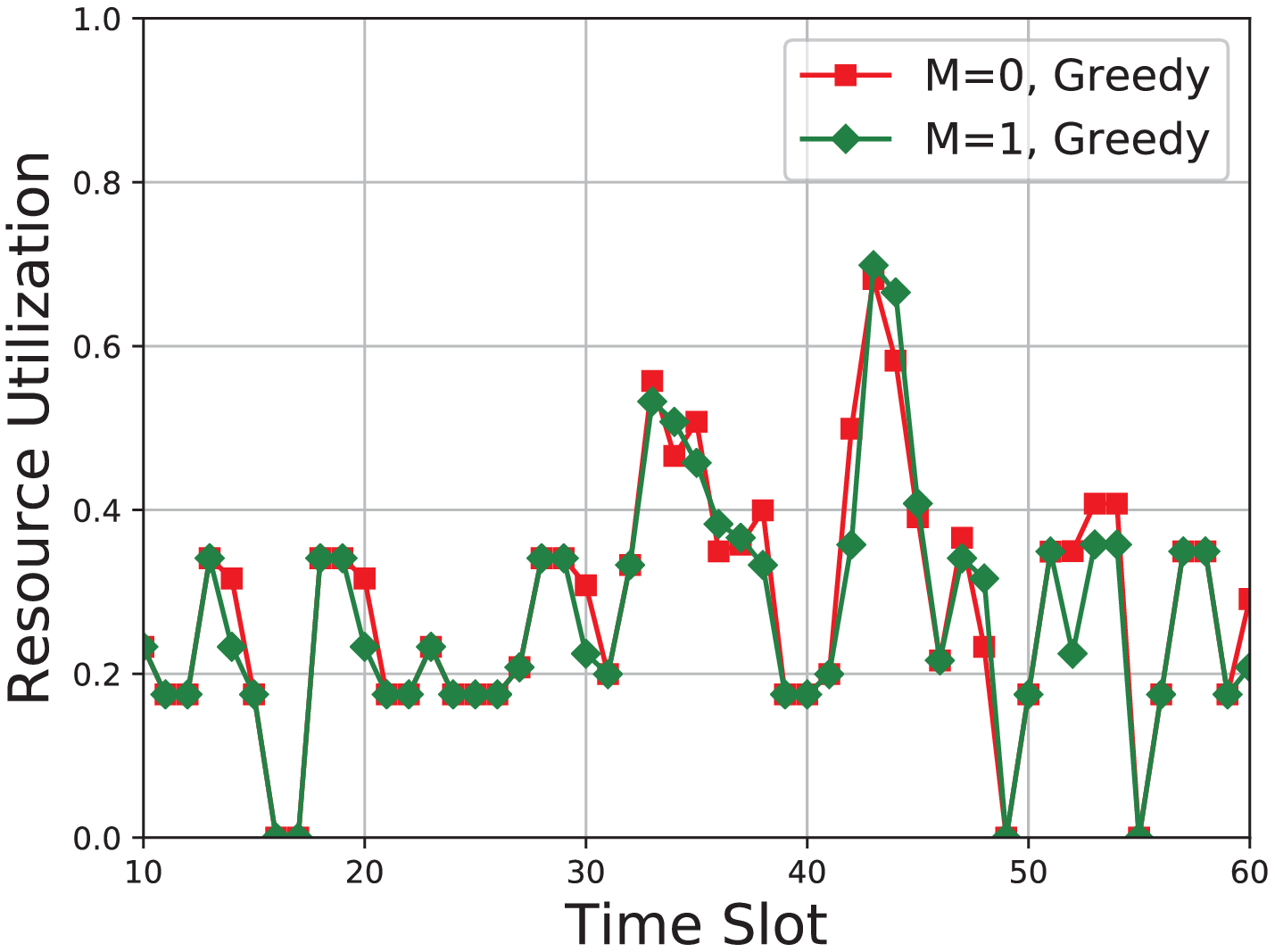}
  \label{LARA algorithm by greedy in BCube network with $4$ servers}}
  \subfigure[Results of LARA algorithms]{\includegraphics[width=0.28\textwidth]{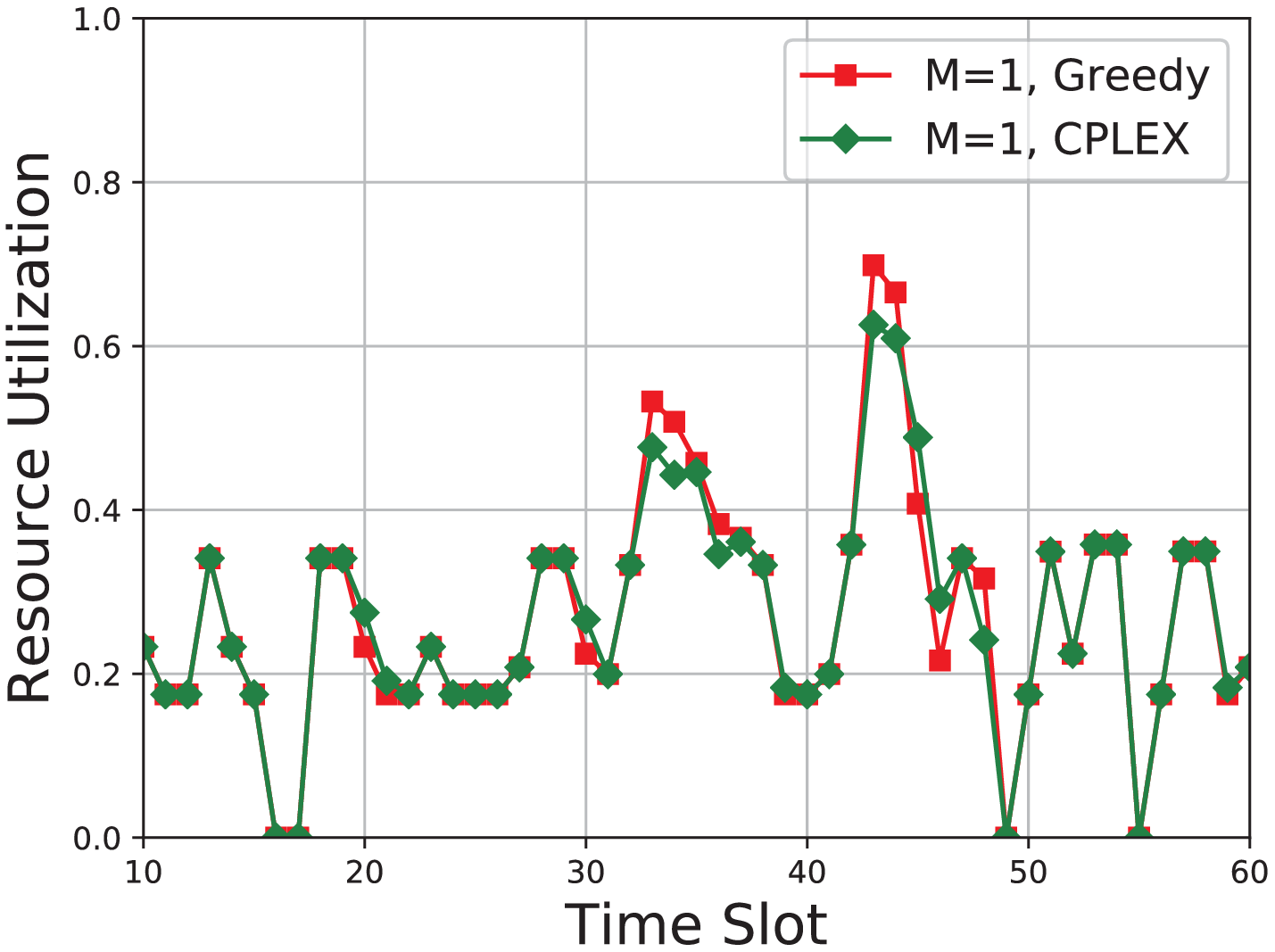}
  \label{Results of LARA algorithms with $4$ servers}}
  \caption{Results of LARA algorithms in a BCube network with $4$ servers.}
  \label{Results of LARA algorithms in BCube network with $4$ servers}
\end{figure*}
\begin{figure*}[tbp]
  \centering
  \subfigure[LARA algorithm by CPLEX]{\includegraphics[width=0.28\textwidth]{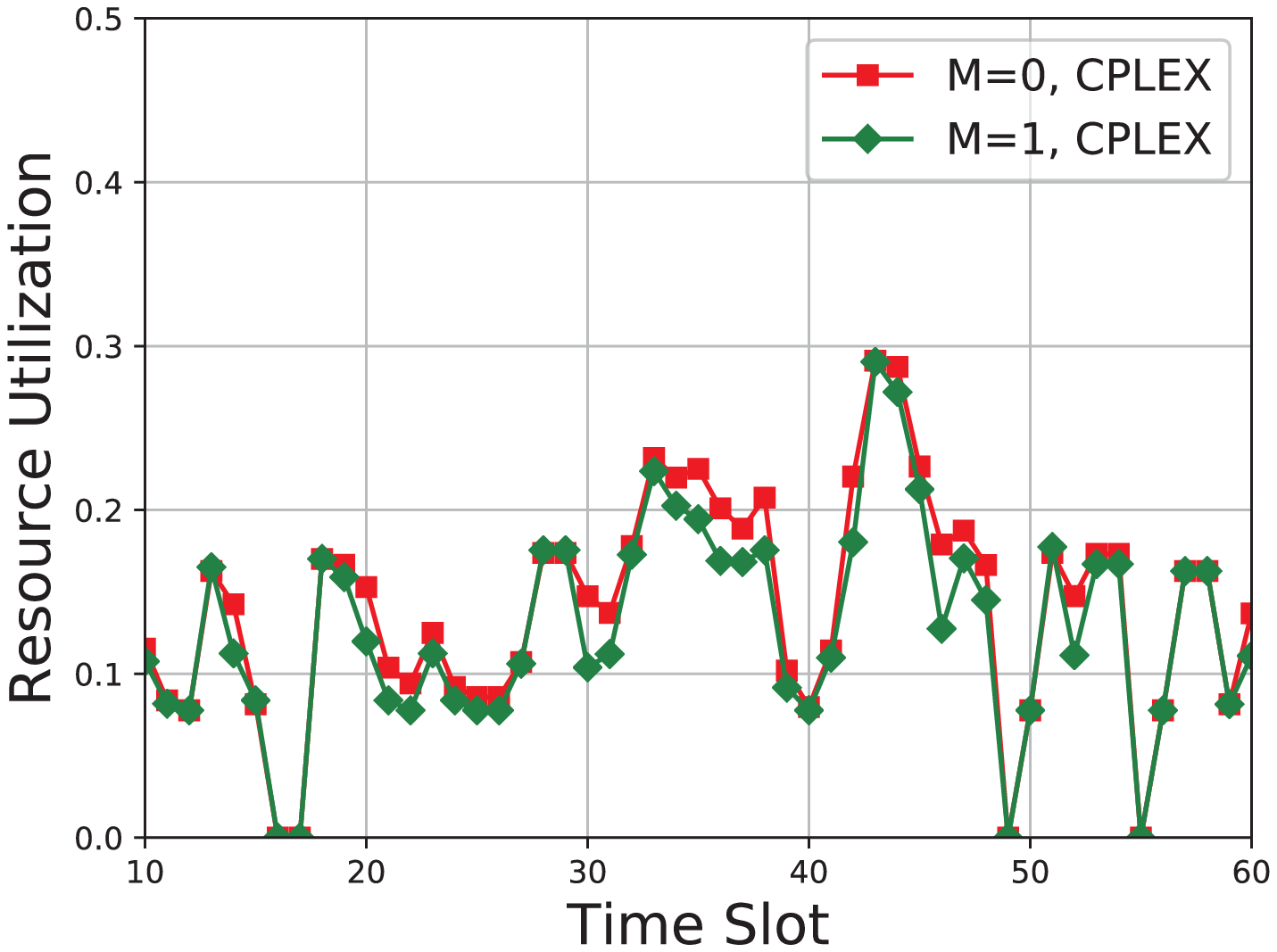}
  \label{LARA algorithm by CPLEX in BCube network with $8$ servers}}
  \subfigure[LARA algorithm by Greedy]{\includegraphics[width=0.28\textwidth]{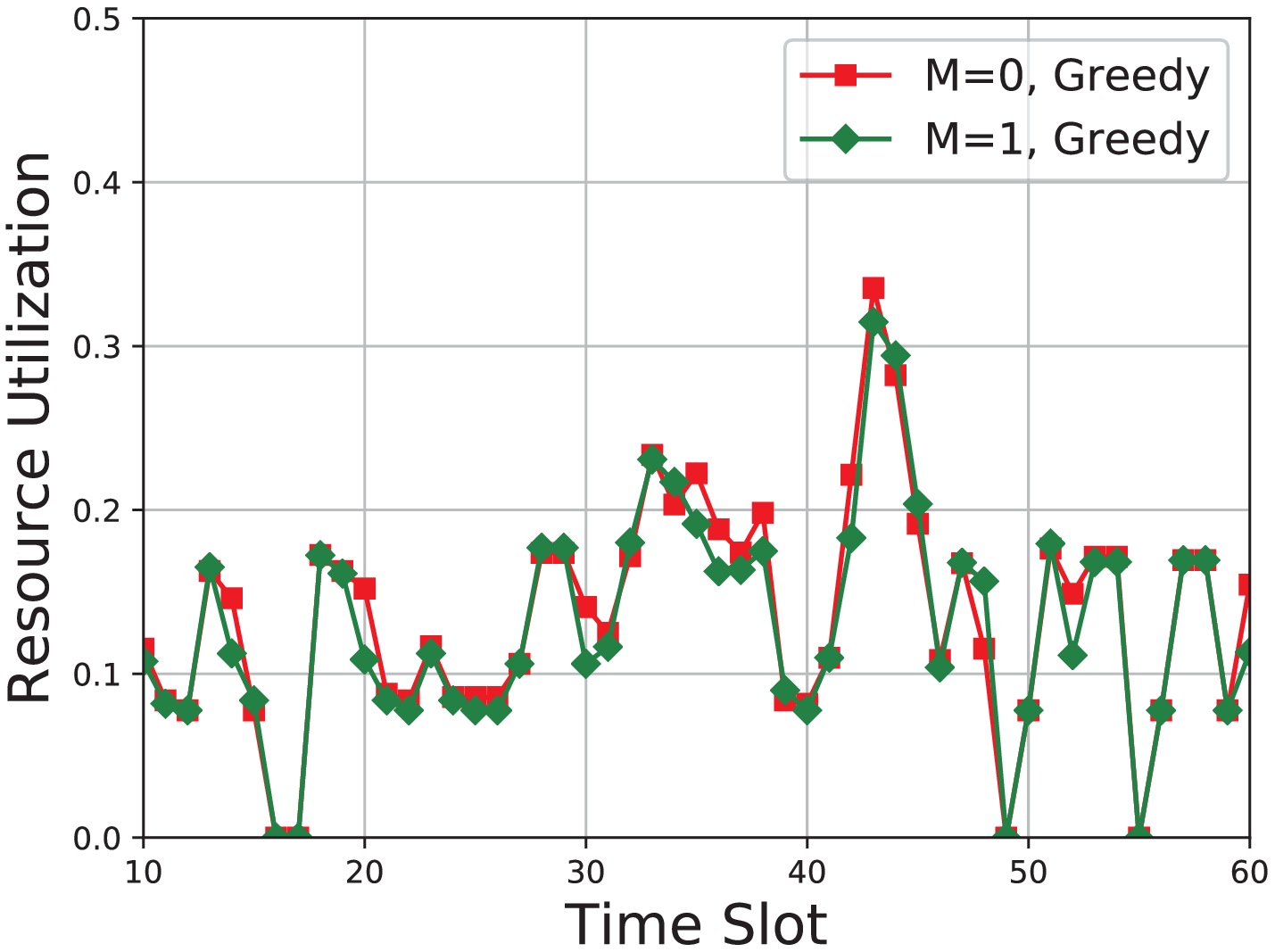}
  \label{LARA algorithm by greedy in BCube network with $8$ servers}}
  \subfigure[Results of LARA algorithms]{\includegraphics[width=0.28\textwidth]{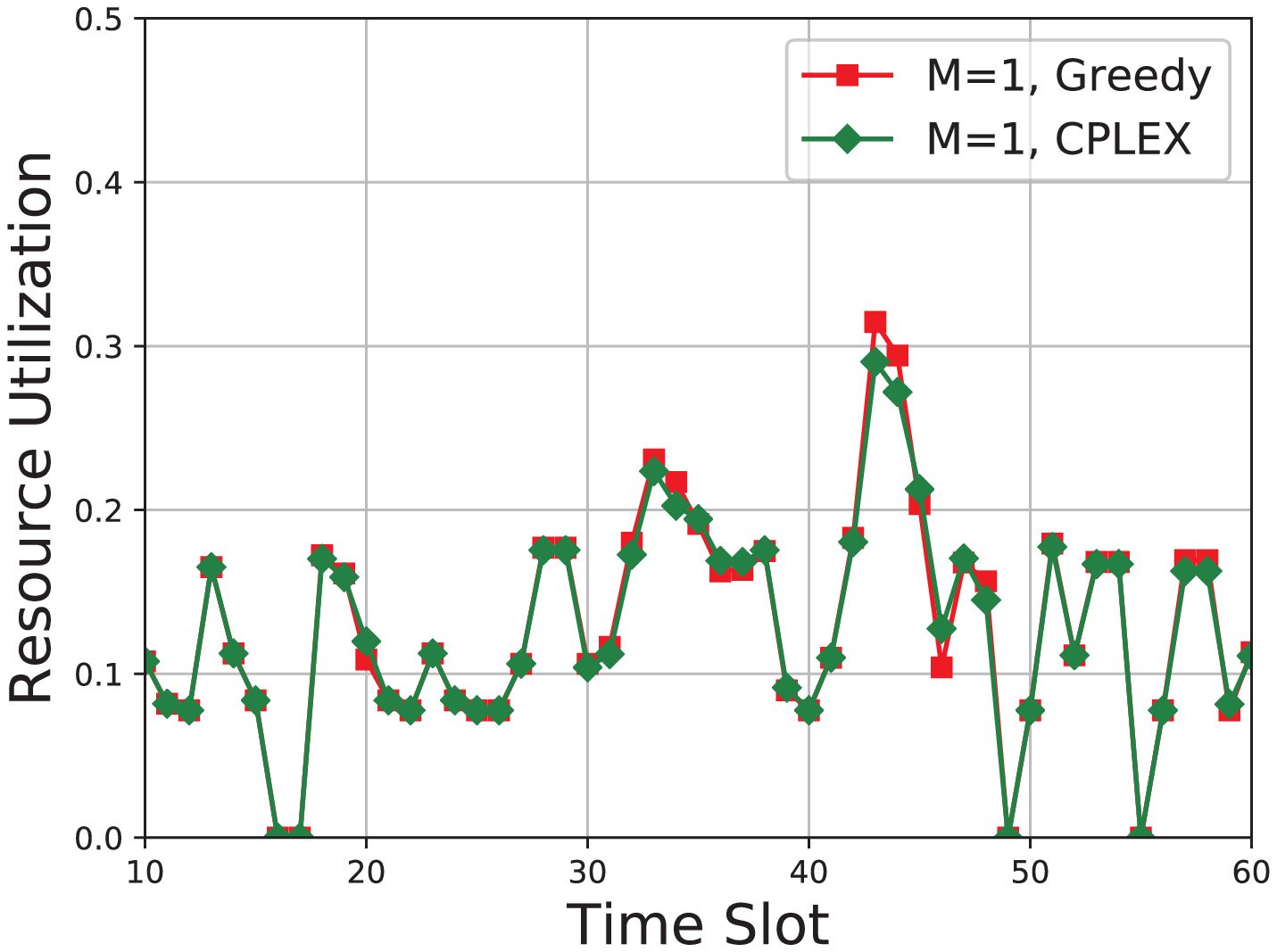}
  \label{Results of LARA algorithms with $8$ servers}}
  \caption{Results of LARA algorithms in a BCube network with $8$ servers.}
  \label{Results of LARA algorithms in BCube network with $8$ servers}
\end{figure*}
\begin{figure*}[tbp]
  \centering
  \subfigure[BCube with $4$ servers]{\includegraphics[width=0.28\textwidth]{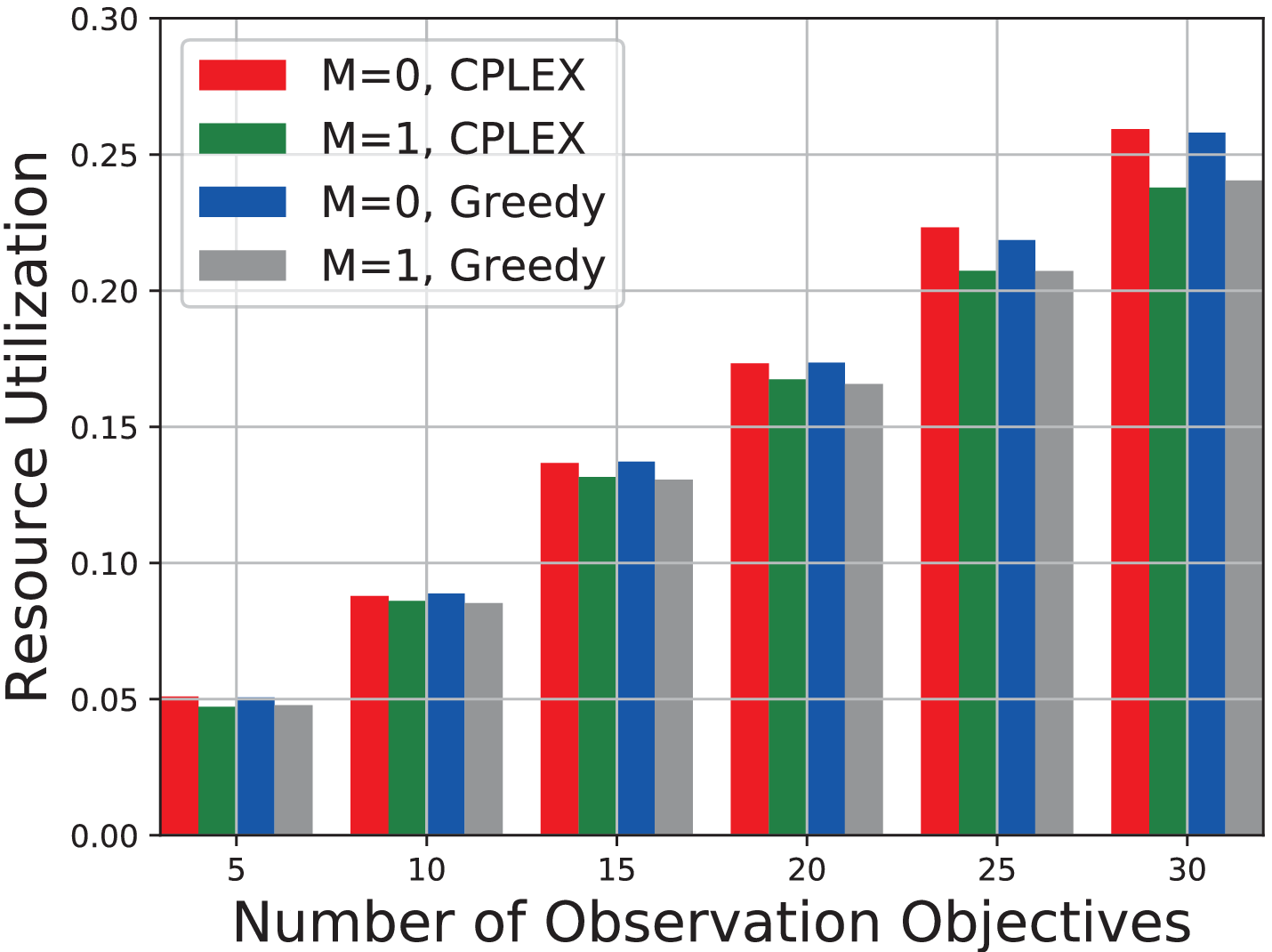}
  \label{BCube with $4$ servers}}
  \subfigure[BCube with $8$ servers]{\includegraphics[width=0.28\textwidth]{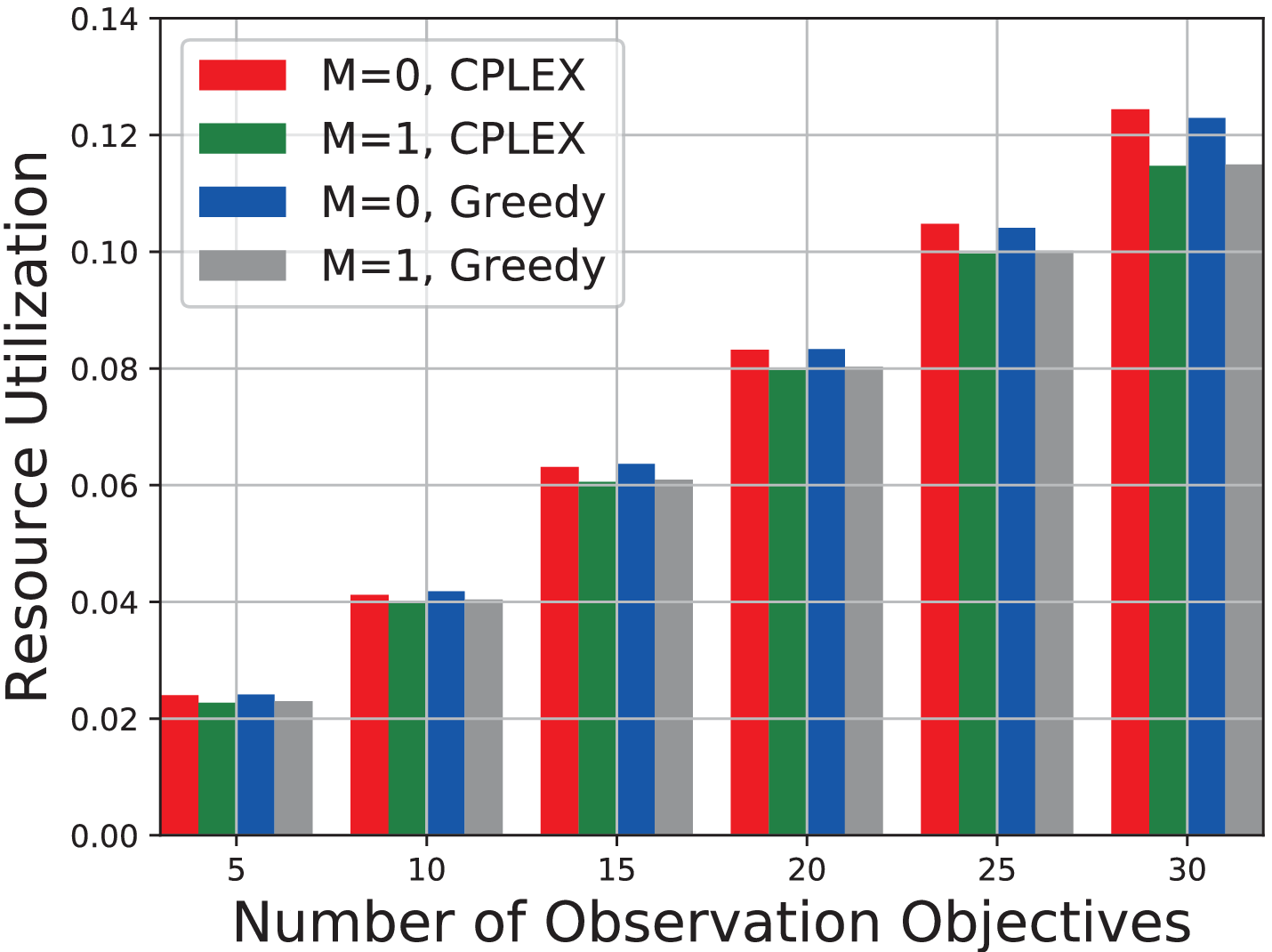}
  \label{BCube with $8$ servers}}
  \subfigure[Time cost]{\includegraphics[width=0.28\textwidth]{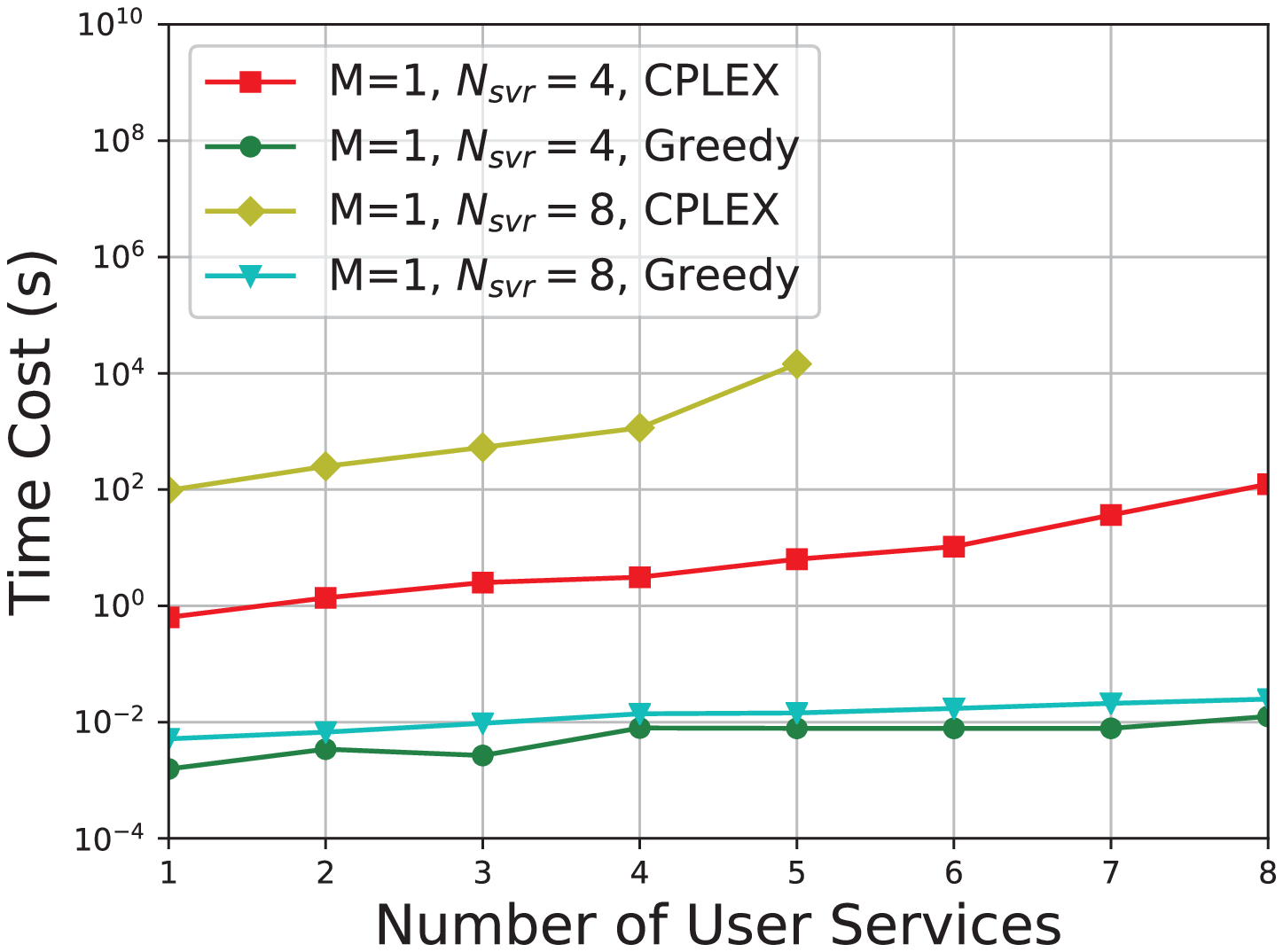}
  \label{Time cost}}
  \caption{Performance comparison between Greedy and CPLEX in BCube networks.}
  \label{Performance comparison between Greedy and CPLEX}
\end{figure*}

\subsection{Performance Comparison of Greedy and CPLEX}\label{Performance Comparison of Greedy and CPLEX}
In this section, we simulate and evaluate the performance of our proposed LARA algorithms based on Greedy and CPLEX in small BCube networks, where the number of servers is $4$ and $8$, respectively. Two situations of predictable and un-predictable user services are taken into consideration in our experiments. Then we discuss the effectiveness of the two proposed LARA algorithms in terms of solution quality and computational cost.\par

Fig.~\ref{Results of LARA algorithms in BCube network with $4$ servers} shows the results of our proposed LARA algorithm in a BCube network with $4$ servers. The number of observation objectives is set as $30$. $M$ indicates the number of predictable time slots, $M=0$ means that the proposed LARA algorithm can not predict the life cycle time of user services. Fig.~\ref{LARA algorithm by CPLEX in BCube network with $4$ servers} and Fig.~\ref{LARA algorithm by greedy in BCube network with $4$ servers} describe the total resource utilizations of the BCube network obtained by the proposed LARA algorithms based on Greedy and CPLEX, respectively. From Fig.~\ref{LARA algorithm by CPLEX in BCube network with $4$ servers}, we can find that our proposed LARA algorithm based on CPLEX for $M=1$ performs better than for $M=0$. From Fig.~\ref{LARA algorithm by greedy in BCube network with $4$ servers}, we can find that our proposed LARA algorithm based on Greedy for $M=1$ also performs better than for $M=0$. Therefore, we can observe that the proposed LARA algorithms with the predictable functionality perform better than the conventional resource allocation algorithms without predictable functionality, i.e., Greedy \cite{liu2015improve} and CPLEX \cite{IBM-CPLEX-WEB}. In Fig.~\ref{Results of LARA algorithms with $4$ servers}, we show the resource utilization results of the proposed LARA algorithms with one predictable time slot. We can observe that the proposed LARA algorithms achieved by Greedy and CPLEX have very similar performance.\par

Similar results are shown for a BCube network with $8$ servers in Fig.~\ref{Results of LARA algorithms in BCube network with $8$ servers}. Fig.~\ref{LARA algorithm by CPLEX in BCube network with $8$ servers} and Fig.~\ref{LARA algorithm by greedy in BCube network with $8$ servers} describe the results of our proposed LARA algorithms based on Greedy and CPLEX, respectively. The performance comparison of our proposed LARA algorithms based on Greedy and CPLEX is illustrated in Fig.~\ref{Results of LARA algorithms with $8$ servers}. Compared with the results as shown in Fig.~\ref{Results of LARA algorithms in BCube network with $4$ servers}, the performance gap between the proposed LARA algorithm and the conventional resource allocation algorithm without predictable functionality could be more obvious in a BCube network with $8$ servers. However, we can observe that our proposed LARA algorithm is an effective approach to improve the performance of solving the problem of VNF placement and routing traffic according to prior sensing the running conditions of user services.\par
\begin{figure*}[tbp]
  \centering
  \subfigure[Fat-Tree network]{\includegraphics[width=0.28\textwidth]{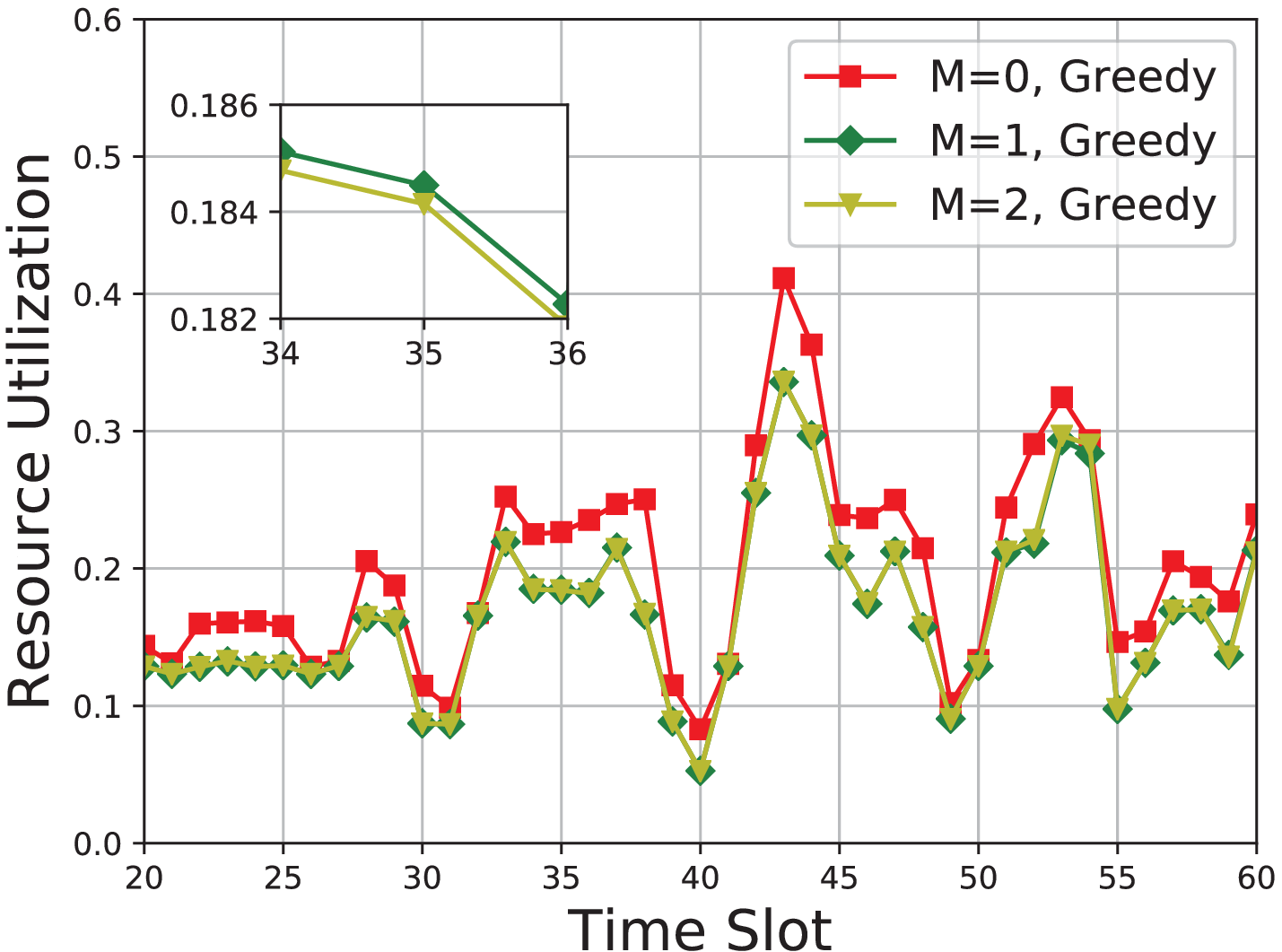}
  \label{Fat-Tree network}}
  \subfigure[BCube network]{\includegraphics[width=0.28\textwidth]{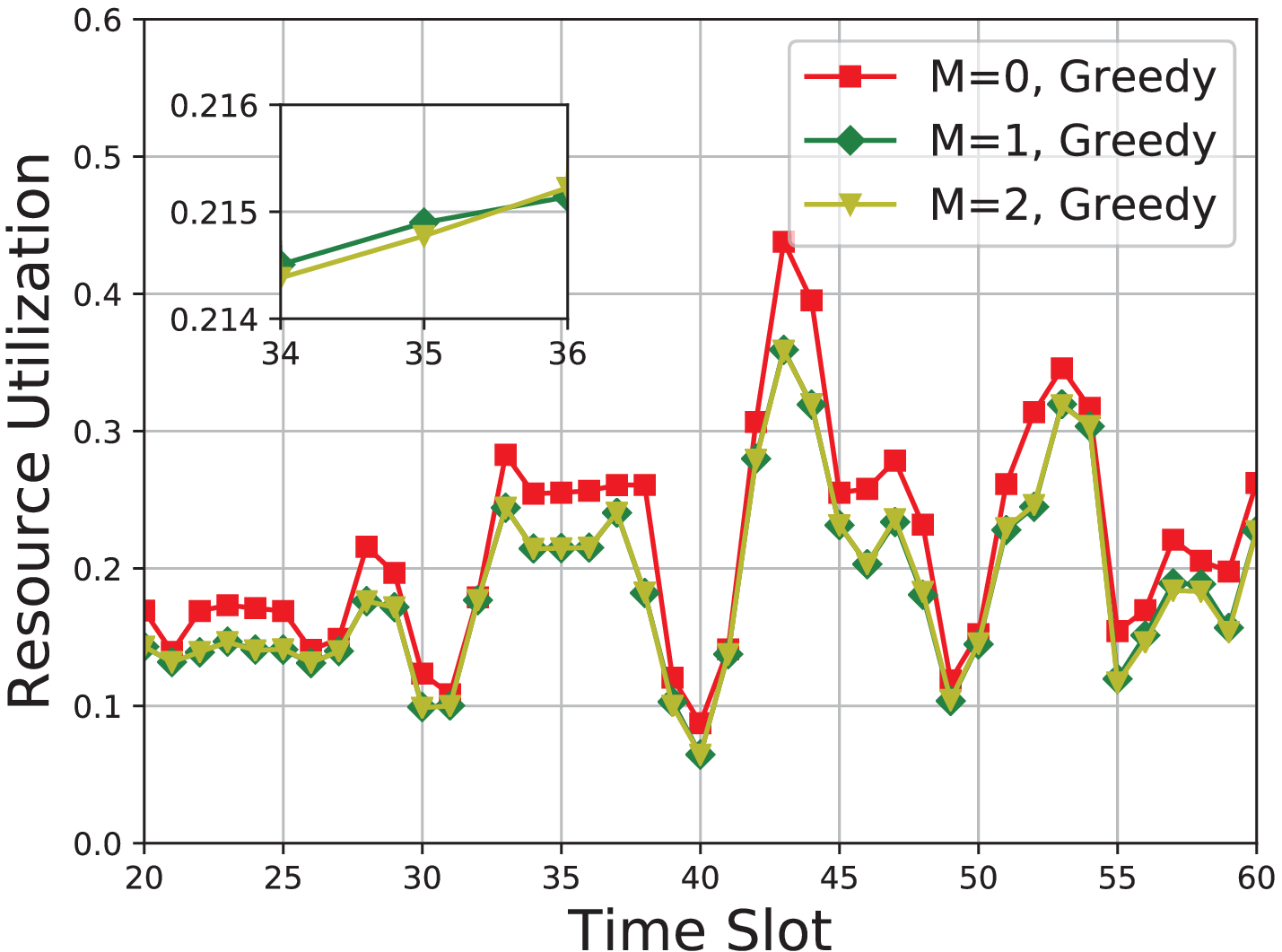}
  \label{BCube network}}
  \subfigure[LV2 network]{\includegraphics[width=0.28\textwidth]{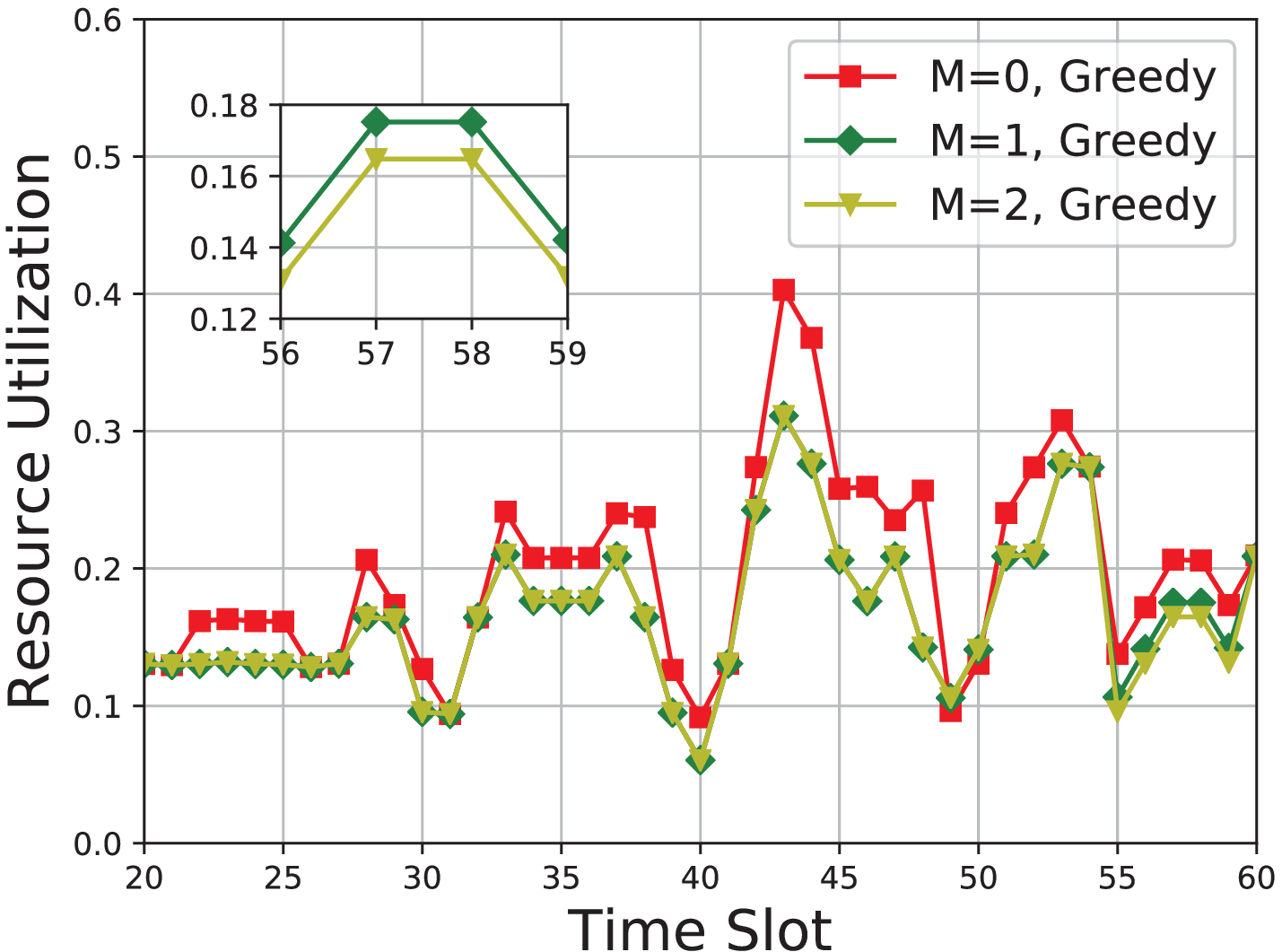}
  \label{LV2 network}}
  \caption{Resource utilizations for Fat-Tree, BCube, and LV2.}
  \label{Resource utilizations for Fat-Tree, BCube and LV2 for 90 user services}
\end{figure*}
\begin{figure*}[tbp]
  \centering
  \subfigure[Resource utilization for servers]{\includegraphics[width=0.28\textwidth]{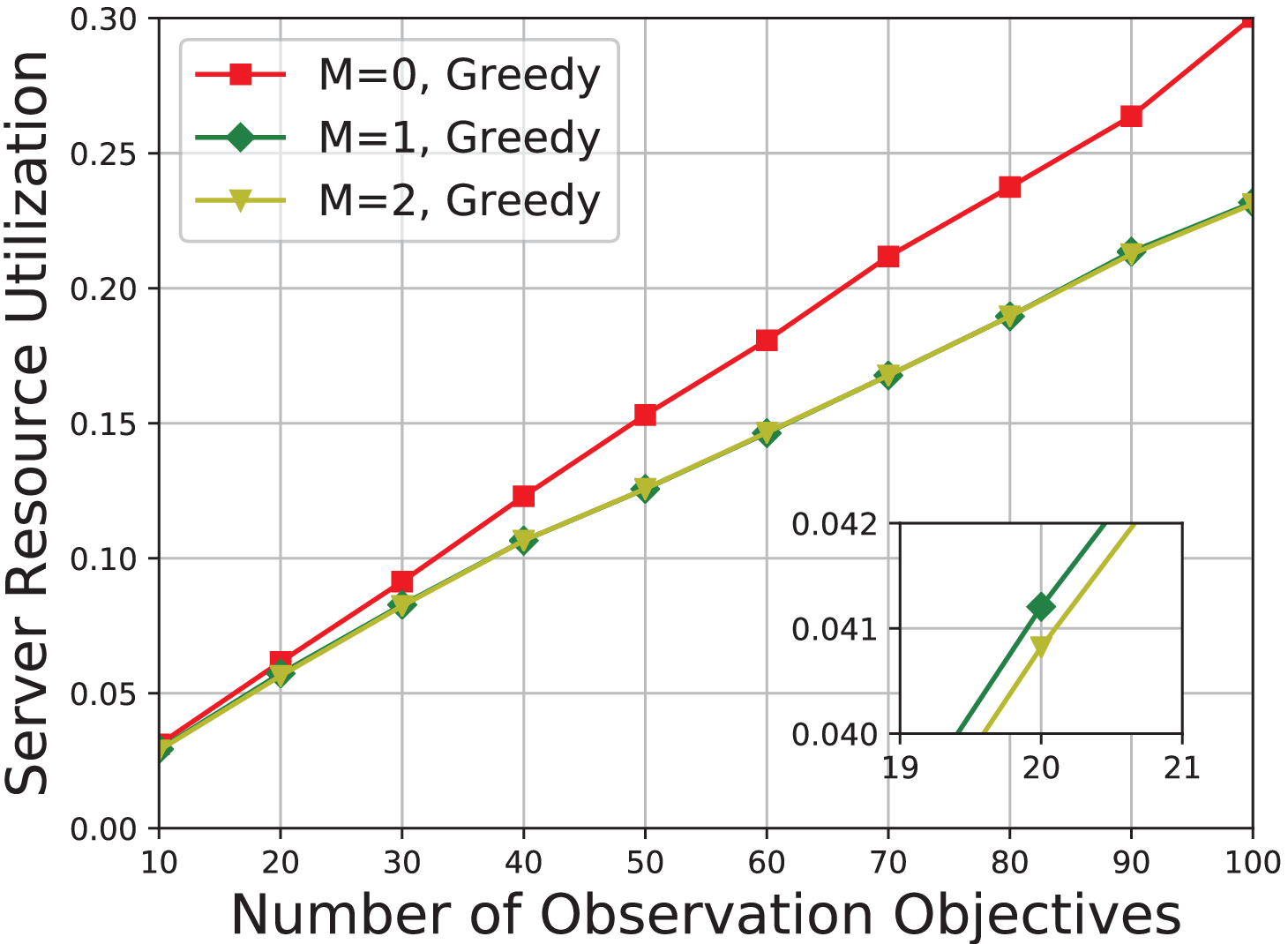}
  \label{Resource utilization for servers}}
  \subfigure[Resource utilization for links]{\includegraphics[width=0.28\textwidth]{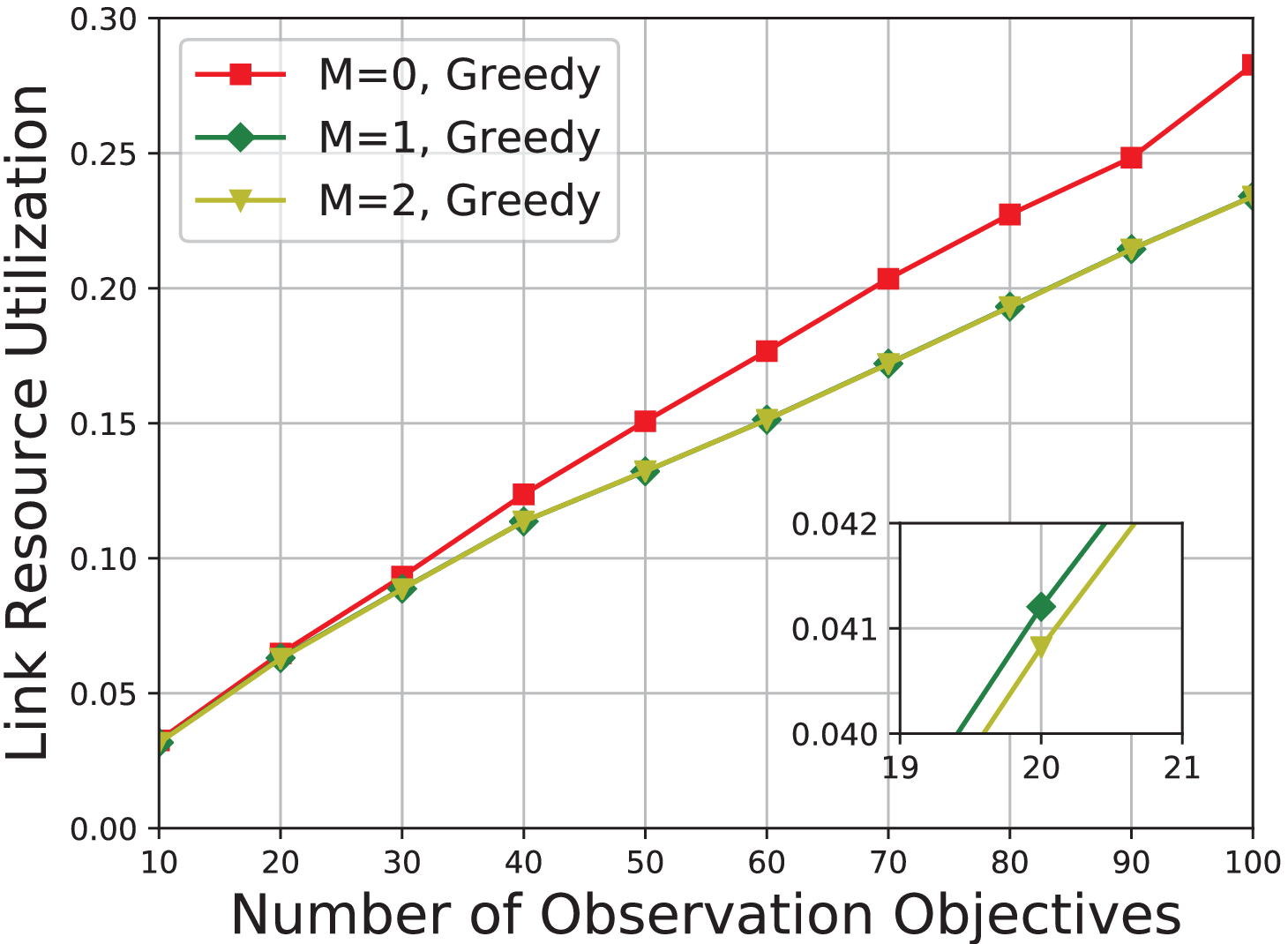}
  \label{Resource utilization for links}}
  \subfigure[Resource utilization]{\includegraphics[width=0.28\textwidth]{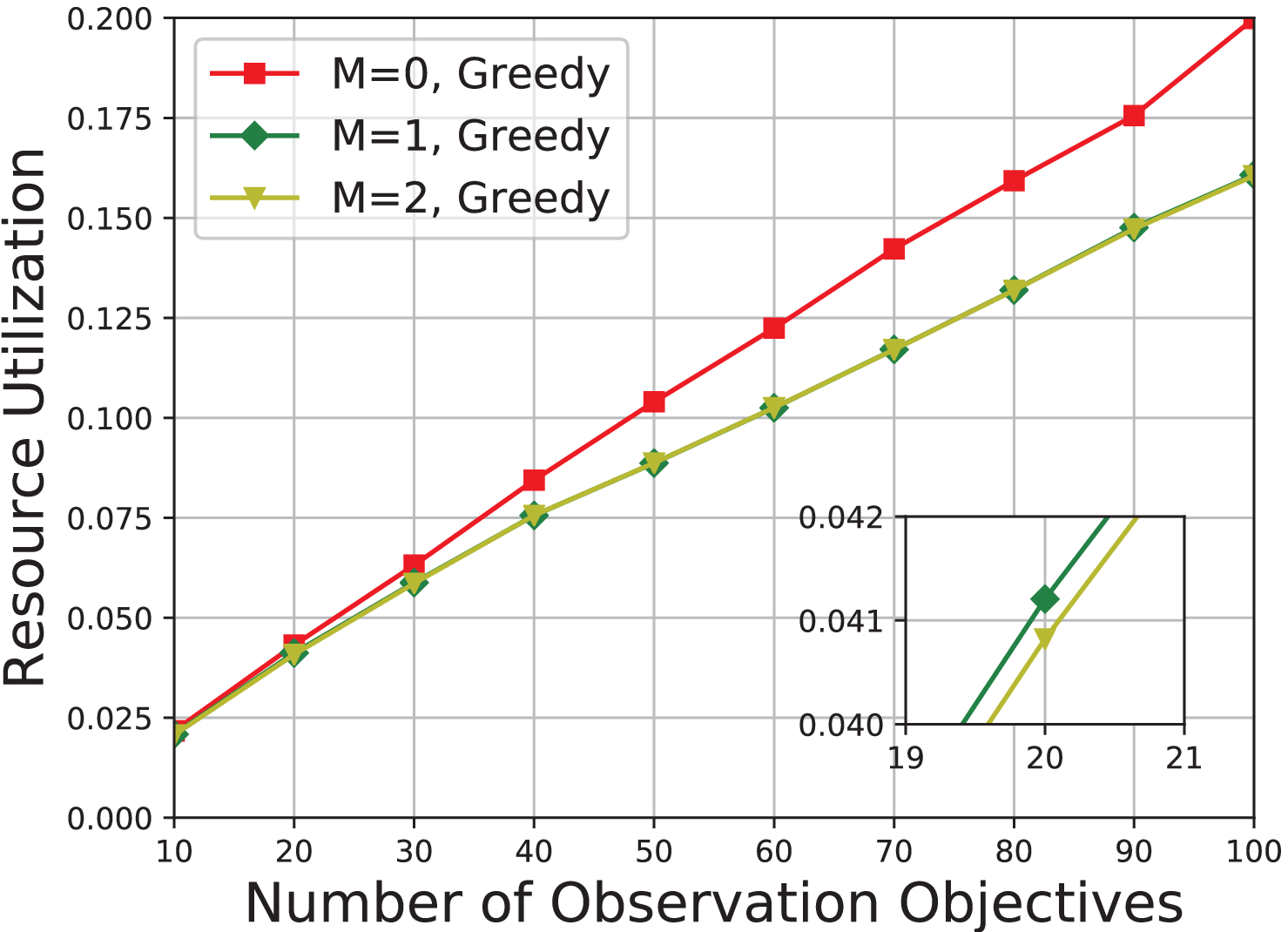}
  \label{Resource utilization}}
  \caption{Resource utilizations for a Fat-Tree network with 16 servers.}
  \label{Resource utilizations for Fat-Tree with 16 servers}
\end{figure*}

In addition, our experiments for different number of observation objectives are carried out to evaluate the performance of the proposed LARA algorithm. The average resource utilizations per time slot for various observation objectives are shown in Fig.~\ref{Performance comparison between Greedy and CPLEX}. The number of observation objectives is denoted as $[5,10,15,20,25,30]$ and the running time for each experiment is $24$ hours. The results of average resource utilizations in BCube networks with $4$ and $8$ servers are depicted in Fig.~\ref{BCube with $4$ servers} and Fig.~\ref{BCube with $8$ servers}, respectively. In all cases of resource allocation, we can find from Fig.~\ref{BCube with $4$ servers} and Fig.~\ref{BCube with $8$ servers} that the performance of our proposed LARA algorithm is better than that of the conventional resource allocation algorithm without predictable functionality. Furthermore, the two proposed LARA algorithms based on Greedy and CPLEX show close results in seeking the solution of resource allocation. For example, the resource utilizations obtained by the proposed LARA algorithms based on Greedy and CPLEX are $0.2404$ and $0.2378$ for $N_{svr}=4,K_{obj}=30,M=1$, and $0.1149$ and $0.1147$ for $N_{svr}=8,K_{obj}=30,M=1$, respectively.\par

The computational time costs for the proposed LARA algorithms based on CPLEX and Greedy are described in Fig.~\ref{Time cost}. Here we consider that the number of user services per time slot is $[1,2,3,4,5,6,7,8]$ due to the computational complexity of CPLEX. BCube networks consist of $4$ and $8$ servers, respectively. We can find that the proposed LARA algorithm based on CPLEX has a long running time for addressing the problem of VNF placement and routing traffic, especially, with the increase in scale of network and number of user services. However, our proposed Greedy-based LARA algorithm can quickly obtain an approximated solution for solving the problem of resource allocation. In a BCube network with $4$ servers, when there are $4$ user services, the average time cost is $3.1077$ seconds for CPLEX and $0.0079$ seconds for Greedy. When the number of user services is $8$, the average time cost is $123.8392$ seconds for CPLEX and $0.0125$ seconds for Greedy. In a BCube network with $8$ servers, when there are $2$ user services, the average time cost is $4.1871$ minutes for CPLEX and $0.0067$ seconds for Greedy. When the number of user services is $5$, the average time cost is $241$ minutes for CPLEX and $0.01427$ seconds for Greedy. We can find that the proposed LARA algorithm based on CPLEX can address the problem of VNF placement and routing traffic in small scale networks, however, it is not suitable to be used in large scale networks. The proposed LARA algorithm based on Greedy in this paper is an effective approach of resource allocation to address the problem of VNF placement and routing traffic in large scale networks.\par

\subsection{Performance Analysis of Greedy-based LARA Algorithm}\label{Performance Analysis}
\begin{table*}[tbp]
  \centering
  \caption{Resource Utilizations for Fat-Tree, BCube, and LV2}
  \label{Resource utilizations for Fat-Tree, BCube and LV2}%
  \resizebox{\textwidth}{!}{
    \begin{tabular}{|p{0.075\textwidth}<{\centering}|p{0.075\textwidth}<{\centering}|p{0.075\textwidth}<{\centering}|p{0.075\textwidth}<{\centering}|p{0.075\textwidth}<{\centering}|p{0.075\textwidth}<{\centering}
    |p{0.075\textwidth}<{\centering}|p{0.075\textwidth}<{\centering}|p{0.075\textwidth}<{\centering}|p{0.075\textwidth}<{\centering}|}
    \hline
    \multirow{2}{*}{$K_{obj}$} & \multicolumn{3}{c|}{Fat-Tree} & \multicolumn{3}{c|}{BCube} & \multicolumn{3}{c|}{LV2} \\
\cline{2-10}          & M=0   & M=1   & M=2   & M=0   & M=1   & M=2   & M=0   & M=1   & M=2 \\
    \hline
    10    & 0.0217 & 0.0209 & 0.0206 & 0.0243  & 0.0236  & 0.0234  & 0.0256 & 0.0247  & 0.0245  \\
    \hline
    20    & 0.0431 & 0.0411 & 0.0408 & 0.0489  & 0.0473  & 0.0471  & 0.0500 & 0.0479  & 0.0475  \\
    \hline
    30    & 0.0634 & 0.0590 & 0.0586 & 0.0714  & 0.0678  & 0.0677  & 0.0704 & 0.0663  & 0.0658  \\
    \hline
    40    & 0.0841 & 0.0754 & 0.0755 & 0.0943  & 0.0869  & 0.0866  & 0.0912 & 0.0827  & 0.0827  \\
    \hline
    50    & 0.1040 & 0.0887 & 0.0887 & 0.1138  & 0.1016  & 0.1020  & 0.1090 & 0.0947  & 0.0947  \\
    \hline
    60    & 0.1222 & 0.1026 & 0.1025 & 0.1343  & 0.1168  & 0.1171  & 0.1264 & 0.1076  & 0.1075  \\
    \hline
    70    & 0.1421 & 0.1173 & 0.1170 & 0.1556  & 0.1329  & 0.1329  & 0.1453 & 0.1205  & 0.1203  \\
    \hline
    80    & 0.1591 & 0.1321 & 0.1319 & 0.1730  & 0.1480  & 0.1482  & 0.1609 & 0.1344  & 0.1340  \\
    \hline
    90    & 0.1758 & 0.1477 & 0.1472 & 0.1905  & 0.1642  & 0.1640  & 0.1753 & 0.1479  & 0.1469  \\
    \hline
    100   & 0.1994 & 0.1606 & 0.1605 & 0.2158  & 0.1789  & 0.1781  & 0.1990 & 0.1594  & 0.1582  \\
    \hline
    \end{tabular}%
    }
\end{table*}%
\begin{figure*}[tbp]
  \centering
  \subfigure[Fat-Tree network with $32$ servers]{\includegraphics[width=0.28\textwidth]{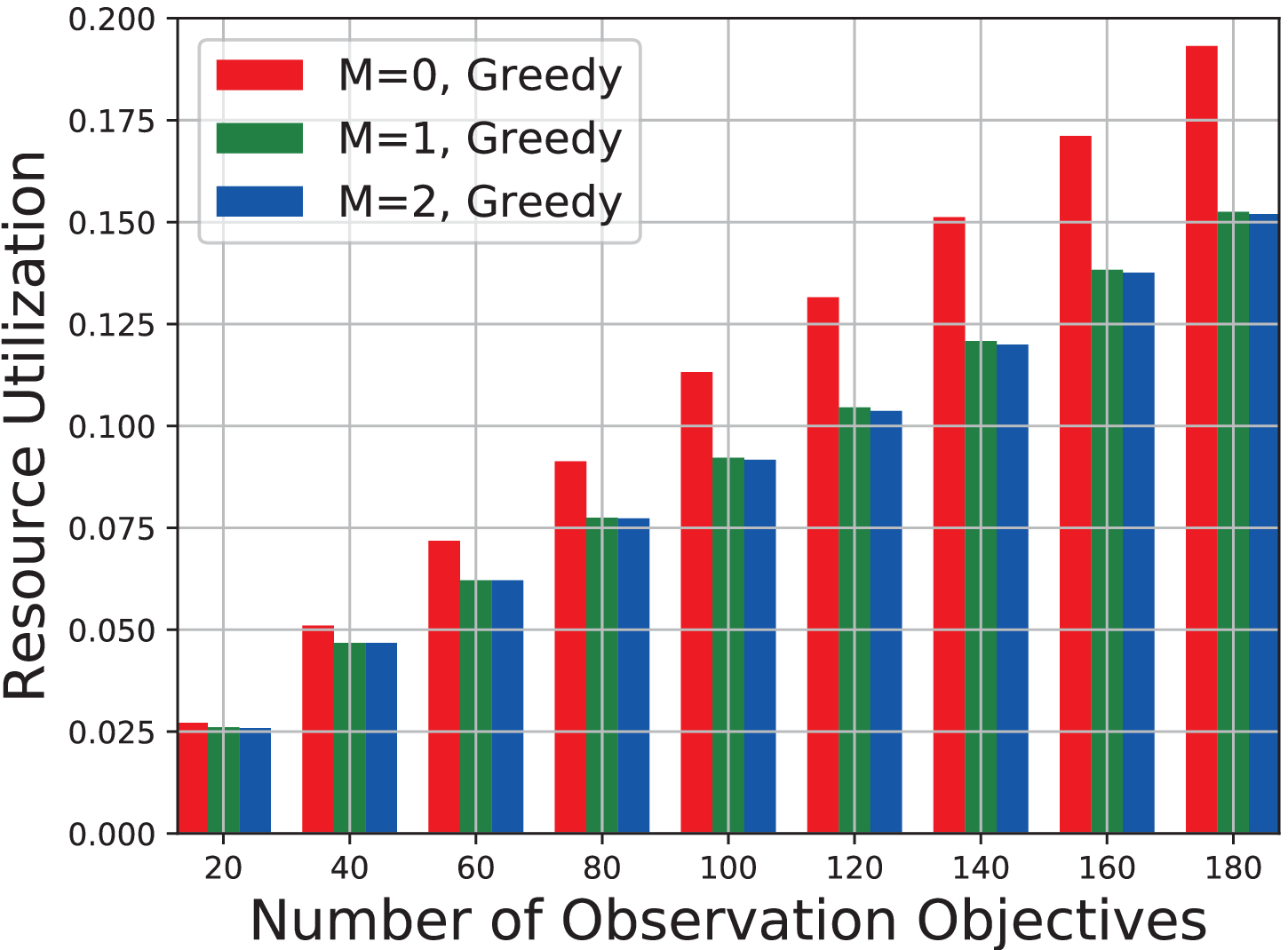}
  \label{Fat-Tree network with 32 servers}}
  \subfigure[Fat-Tree network with $48$ servers]{\includegraphics[width=0.28\textwidth]{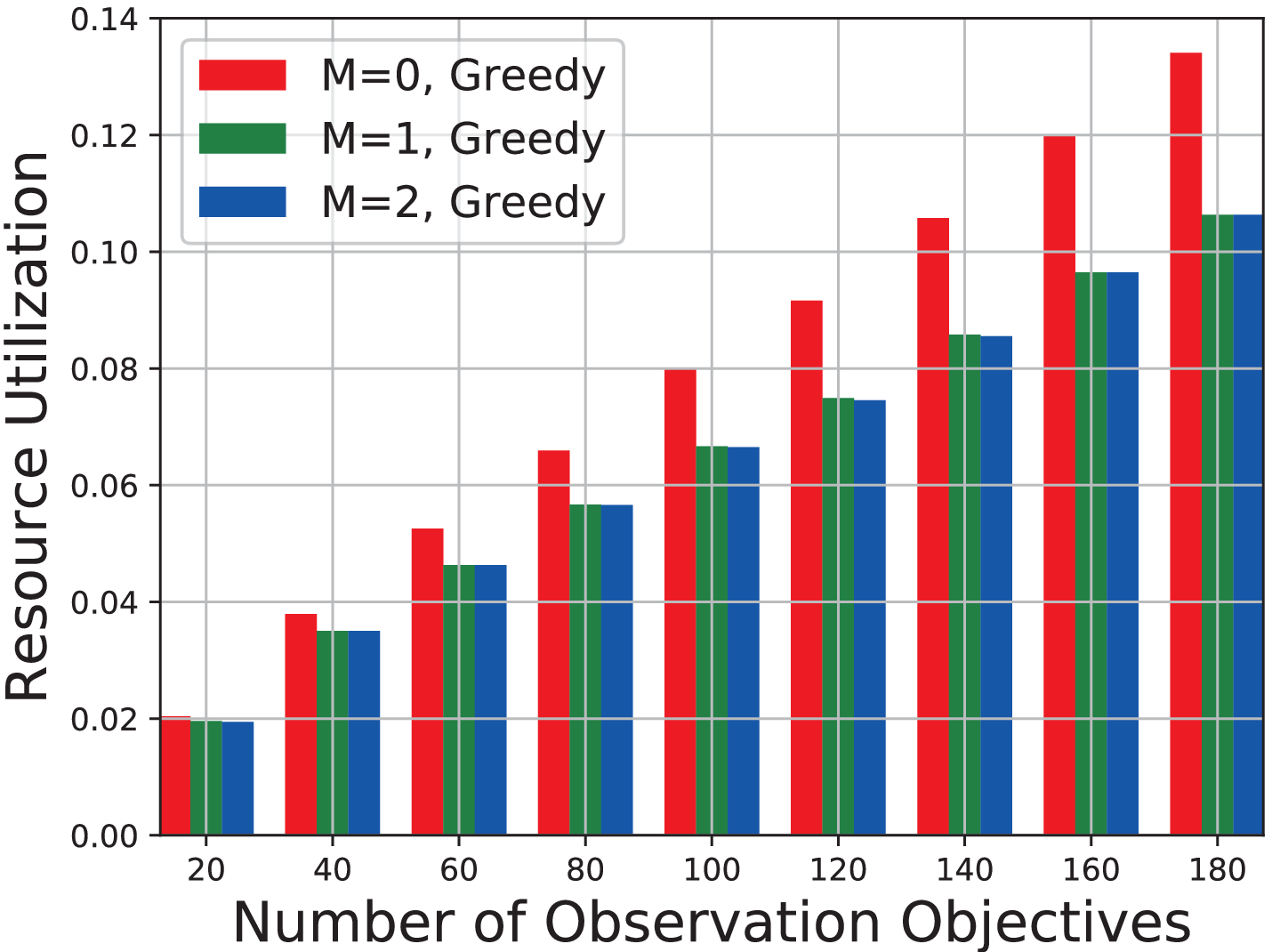}
  \label{Fat-Tree network with 48 servers}}
  \subfigure[Fat-Tree network with $64$ servers]{\includegraphics[width=0.28\textwidth]{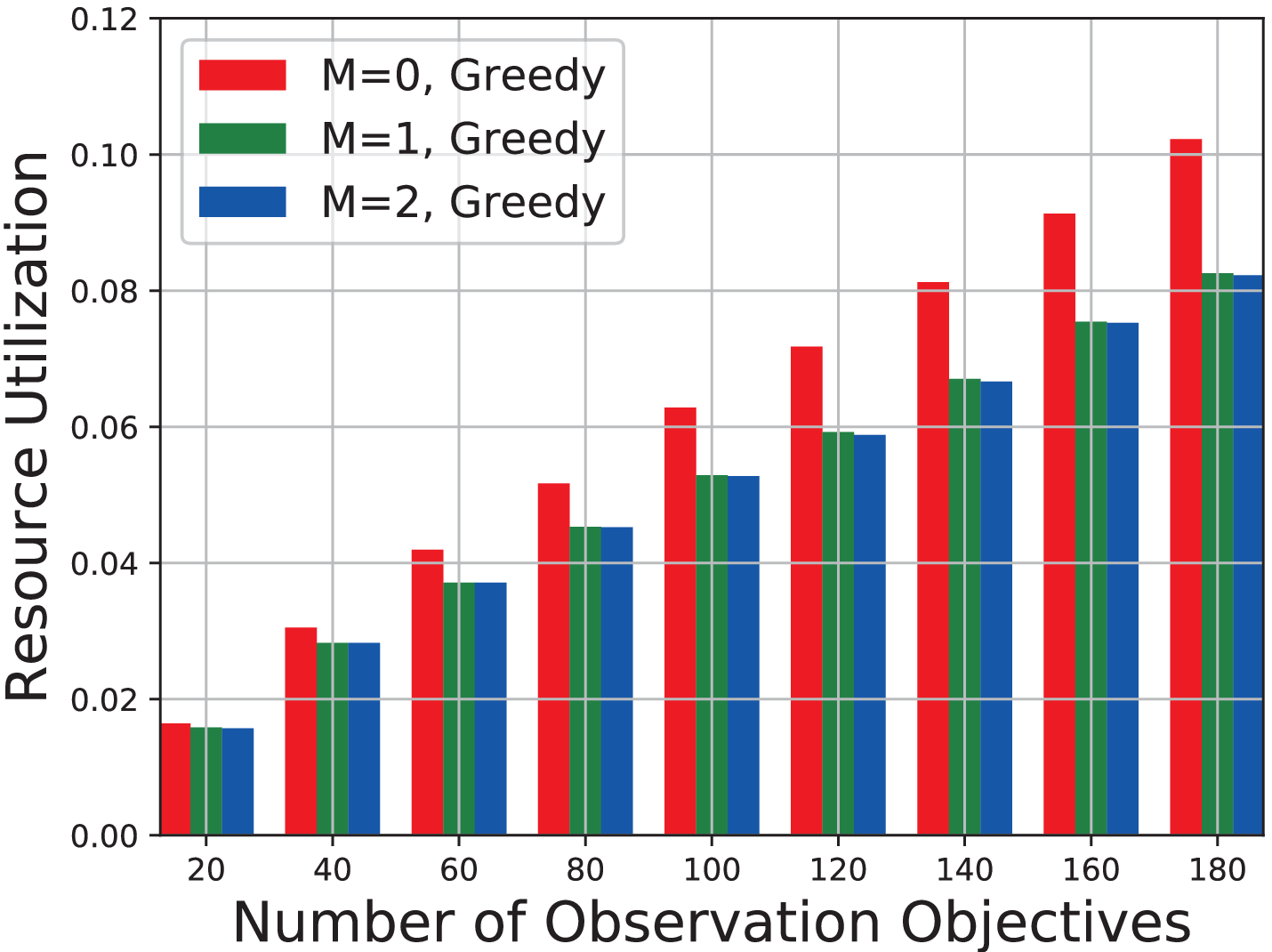}
  \label{Fat-Tree network with 64 servers}}
  \caption{Resource utilizations for Fat-Tree networks with $32, 48$, and $64$ servers.}
  \label{Resource utilizations for Fat-Tree networks with different servers}
\end{figure*}
In this section, we evaluate the performance of the proposed Greedy-based LARA algorithm for multiple predictable time slots in three network structures of Fat-Tree, BCube, and LV2 with $16$ servers, respectively. The number of observation objectives is from $10$ to $100$ and the running time for each experiment is $24$ hours. The predictable time slots are $0,1,$ and $2$, respectively. Each experiment is carried out $50$ times and we obtain the average resource utilization results in terms of servers, bandwidths, and links.\par

Fig.~\ref{Resource utilizations for Fat-Tree, BCube and LV2 for 90 user services} shows the results of resource utilizations obtained by the proposed Greedy-based LARA algorithm for 90 observation objectives in Fat-Tree, BCube, and LV2 networks, respectively. In Fig.~\ref{Fat-Tree network}, the results of resource utilizations for $M=0,1,$ and $2$ in a Fat-Tree network with 32 servers are illustrated. We can observe that the proposed Greedy-based LARA algorithm performs better than the conventional Greedy-based resource allocation algorithm without predictable functionality, e.g., $M=0$. The performance of the proposed Greedy-based LARA algorithm is relatively close as the number of predictable time slots increases under our simulation parameters setup. The proposed Greedy-based LARA algorithm for $M=2$ performs slightly better than that of $M=1$. Similar results for BCube and LV2 networks can be found in Fig.~\ref{BCube network} and Fig.~\ref{LV2 network}, respectively. It is obvious that our proposed Greedy-based LARA algorithm can effectively decrease the resource utilization of the three networks by introducing the predictable functionality.\par

To further investigate the influence of different number of observation objectives on the performance, we run the experiments for $K_{obj}=[10,20,\cdots,100]$ by the proposed Greedy-based LARA algorithm in a Fat-Tree network with 16 servers and the average results of resource utilizations are shown in Fig.~\ref{Resource utilizations for Fat-Tree with 16 servers}. The proposed Greedy-based LARA algorithm with $M=0$ is considered as our baseline algorithm. Fig.~\ref{Resource utilization for servers} illustrates the resource utilizations of servers for different number of observation objectives. We can observe that the proposed Greedy-based LARA algorithms for $M=0,1,$ and $2$ have relatively close results in the case of the small number of observation objectives, and our proposed Greedy-based LARA algorithm performs better with the increase in the number of observation objectives and predictable time slots, respectively. For instance, in the case of $K_{obj}=90$, the performance improvement of our proposed Greedy-based LARA algorithm in the resource utilization of servers is $19.03\%$ for $M=1$ and $19.38\%$ for $M=2$. On average, the resource utilization of servers obtained by the proposed Greedy-based LARA algorithm reduces by $18.35\%$ for $M=1$ and $18.53\%$ for $M=2$. The resource utilizations of links for different number of observation objectives are shown in Fig.~\ref{Resource utilization for links}. We can observe that our proposed Greedy-based LARA algorithm effectively decreases the number of used links in assigning network resources for user services. For $K_{obj}=90$, the resource utilization of links obtained by our proposed Greedy-based LARA algorithm reduces by $13.63\%$ for $M=1$ and $13.67\%$ for $M=2$. On average, the link resource utilization of our proposed Greedy-based LARA algorithm saves by $13.03\%$ for $M=1$ and $13.09\%$ for $M=2$. The total resource utilizations for different observation objectives are described in Fig.~\ref{Resource utilization}. As shown in Fig.~\ref{Resource utilization}, the average resource utilization gained by our proposed Greedy-based LARA algorithm decreases by $15.20\%$ for $M=1$ and $15.34\%$ for $M=2$, respectively.\par

In order to evaluate the performance of our proposed Greedy-based LARA algorithm in three network structures of Fat-Tree, BCube, and LV2, we make the experiments for different observation objectives in Fat-Tree, BCube, and LV2. Each experiment is carried out $50$ times and the average results of resource utilizations are shown in Table \ref{Resource utilizations for Fat-Tree, BCube and LV2}. We can observe that our proposed Greedy-based LARA algorithm performs better than the Greedy-based resource allocation algorithm without predictable functionality in the three networks. For $M=1$, the average resource utilizations obtained by our proposed Greedy-based LARA algorithm for Fat-Tree, BCube, and LV2 decrease by $15.20\%,12.59\%,$ and $14.48\%$, respectively. In the case of $M=2$, our proposed Greedy-based LARA algorithm for Fat-Tree, BCube, and LV2 has $15.34\%,12.62\%,$ and $14.78\%$ performance improvement on average, respectively. Hence it can be stated that our proposed Greedy-based LARA algorithm can effectively improve the performance of resource utilizations for the three networks of Fat-Tree, BCube, and LV2.\par

Furthermore, to evaluate the performance of the proposed LARA algorithm based on Greedy as the number of servers increases, we make the following experiments in Fat-Tree networks with $32,48$, and $64$, respectively. The number of observation objectives is $[20,40,60,80,100,120,140,160,180]$. The number of predictable time slots is $0,1$, and $2$, respectively. Each experiment is repeated $50$ times and the running time for each experiment is $24$ hours. Then we obtain the average results of resource utilizations. Fig.~\ref{Resource utilizations for Fat-Tree networks with different servers} shows that the average resource utilizations for different number of observation objectives in Fat-Tree networks with $32,48$, and $64$ servers, respectively. The average resource utilizations for different number of observation objectives in a Fat-Tree network with $32$ servers are shown in Fig.~\ref{Fat-Tree network with 32 servers}. We can observe from Fig.~\ref{Fat-Tree network with 32 servers} that the resource utilization results for $M=0,1$, and $2$ are relatively close when the number of observation objectives is small, e.g., $K_{obj}=20$. With the increase in the number of observation objectives, the proposed Greedy-based LARA algorithm performs better than the conventional Greedy-based resource allocation algorithm, for $M=0$ case. The proposed Greedy-based LARA algorithm for $M=2$ performs slightly better than the case of $M=1$. For example, when $K_{obj}=100$, the resource utilizations for $M=0,1$, and $2$ are $0.1132,0.0921$, and $0.0916$, respectively. Compared with the baseline resource allocation algorithm with $M=0$, the resource utilization obtained by the proposed Greedy-based LARA algorithm reduces by $18.58\%$ for $M=1$ and $19.01\%$ for $M=2$. On average, the performance improvement of the proposed Greedy-based LARA algorithm is $18.05\%$ for $M=1$ and $18.43\%$ for $M=2$ in a Fat-Tree network with $32$ servers, respectively. Similar results for Fat-Tree networks with $48$ and $64$ servers are shown in Fig.~\ref{Fat-Tree network with 48 servers} and Fig.~\ref{Fat-Tree network with 64 servers}. For a Fat-Tree network with $48$ servers, the average resource utilization obtained by the proposed Greedy-based LARA algorithm decreases by $17.01\%$ for $M=1$ and $17.16\%$ for $M=2$, respectively. For a Fat-Tree network with $64$ servers, the average resource utilization obtained by the proposed Greedy-based LARA algorithm decreases by $15.71\%$ for $M=1$ and $15.99\%$ for $M=2$, respectively. We can observe from Fig.~\ref{Resource utilizations for Fat-Tree networks with different servers} that the proposed Greedy-based LARA algorithm is effective for service chaining placement in satellite ground station networks when the number of servers increases.\par
\section{Conclusion}\label{Conclusion}

In this paper, considering that the information about service types, resource requirements, and the running time for user services can be known beforehand depending on satellite mission planning in satellite control centers, we investigate the problem of service chaining placement in satellite ground station networks. We formulate the problem of VNF placement and routing traffic as an integer linear programming model and prove it as NP-hard. Our goal is to minimize the resource utilization of the underlying network within the physical resource constraints.\par

To address this problem, The LARA algorithms based on Greedy and CPLEX are implemented. We simulate and evaluate the performance of the two proposed LARA algorithms in small scale networks of BCube with 4 and 8 servers, respectively. The results show that the proposed LARA algorithms based on CPLEX and Greedy have close performance, where the CPLEX-based LARA algorithm can be used in small scale networks due to the computational complexity. To further discuss the performance of our proposed LARA algorithm, we use the proposed Greedy-based LARA algorithm to address the problem of resource allocation in three networks of Fat-Tree, BCube, and LV2 with 16 servers, respectively. We can find that our proposed Greedy-based LARA algorithm outperforms the Greedy-based resource allocation algorithm for the three networks in the resource utilizations of SGS networks. In addition, the number of predictable time slots has a slight effect on the performance of our proposed LARA algorithm. The resource utilizations of Fat-Tree, BCube, and LV2 obtained by our proposed Greedy-based LARA algorithm can decrease by $15.20\%,12.59\%,$ and $14.48\%$ for $M=1$, and $15.34\%,12.62\%,$ and $14.78\%$ for $M=2$ on average. We also evaluate the performance of the proposed Greedy-based LARA algorithm in Fat-Tree networks as the number of servers increases. The simulation results demonstrate the effectiveness of the proposed Greedy-based LARA algorithm with the increase in the number of servers.


%

%

%
%

\ifCLASSOPTIONcaptionsoff
  \newpage
\fi



\bibliographystyle{IEEEtran}
\bibliography{Service_Chaining_Placement_Based_on_Satellite_Mission_Planning_in_Ground_Station_Networks}

\begin{thebibliography}{10}
\providecommand{\url}[1]{#1}
\csname url@samestyle\endcsname
\providecommand{\newblock}{\relax}
\providecommand{\bibinfo}[2]{#2}
\providecommand{\BIBentrySTDinterwordspacing}{\spaceskip=0pt\relax}
\providecommand{\BIBentryALTinterwordstretchfactor}{4}
\providecommand{\BIBentryALTinterwordspacing}{\spaceskip=\fontdimen2\font plus
\BIBentryALTinterwordstretchfactor\fontdimen3\font minus
  \fontdimen4\font\relax}
\providecommand{\BIBforeignlanguage}[2]{{%
\expandafter\ifx\csname l@#1\endcsname\relax
\typeout{** WARNING: IEEEtran.bst: No hyphenation pattern has been}%
\typeout{** loaded for the language `#1'. Using the pattern for}%
\typeout{** the default language instead.}%
\else
\language=\csname l@#1\endcsname
\fi
#2}}
\providecommand{\BIBdecl}{\relax}
\BIBdecl

\bibitem{8066287}
J.~H. {Cox}, J.~{Chung}, S.~{Donovan} \emph{et~al.}, ``Advancing
  software-defined networks: A survey,'' \emph{IEEE Access}, vol.~5, pp.
  25\,487--25\,526, 2017.

\bibitem{7243304}
R.~{Mijumbi}, J.~{Serrat}, J.~{Gorricho} \emph{et~al.}, ``Network function
  virtualization: State-of-the-art and research challenges,'' \emph{IEEE
  Commun. Surveys Tuts.}, vol.~18, no.~1, pp. 236--262, 2016.

\bibitem{medhat2017service}
A.~M. Medhat, T.~Taleb, A.~Elmangoush \emph{et~al.}, ``Service function
  chaining in next generation networks: State of the art and research
  challenges,'' \emph{IEEE Commun. Mag.}, vol.~55, no.~2, pp. 216--223, 2017.

\bibitem{kar2018energy}
B.~Kar, E.~H. Wu, and Y.~Lin, ``Energy cost optimization in dynamic placement
  of virtualized network function chains,'' \emph{IEEE Trans. Netw. Serv.
  Manag.}, vol.~15, no.~1, pp. 372--386, 2018.

\bibitem{herrera2016resource}
J.~G. Herrera and J.~F. Botero, ``Resource allocation in {NFV}: A comprehensive
  survey,'' \emph{IEEE Trans. Netw. Serv. Manag.}, vol.~13, no.~3, pp.
  518--532, 2016.

\bibitem{bhamare2016a}
D.~Bhamare, R.~Jain, M.~Samaka \emph{et~al.}, ``A survey on service function
  chaining,'' \emph{J. Netw. Comput. Appl.}, vol.~75, pp. 138--155, 2016.

\bibitem{rankothge2017optimizing}
W.~Rankothge, F.~Le, A.~Russo \emph{et~al.}, ``Optimizing resource allocation
  for virtualized network functions in a cloud center using genetic
  algorithms,'' \emph{IEEE Trans. Netw. Serv. Manag.}, vol.~14, no.~2, pp.
  343--356, 2017.

\bibitem{bari2016orchestrating}
F.~Bari, S.~R. Chowdhury, R.~Ahmed \emph{et~al.}, ``Orchestrating virtualized
  network functions,'' \emph{IEEE Trans. Netw. Serv. Manag.}, vol.~13, no.~4,
  pp. 725--739, 2016.

\bibitem{8060513}
F.~Z. {Yousaf}, M.~{Bredel}, S.~{Schaller} \emph{et~al.}, ``{NFV} and {SDN}-key
  technology enablers for {5G} networks,'' \emph{IEEE J. Sel. Area. Commun.},
  vol.~35, no.~11, pp. 2468--2478, 2017.

\bibitem{8685771}
B.~{Feng}, G.~{Li}, G.~{Li} \emph{et~al.}, ``Enabling efficient service
  function chains at terrestrial-satellite hybrid cloud networks,'' \emph{IEEE
  Netw.}, vol.~33, no.~6, pp. 94--99, 2019.

\bibitem{7962772}
T.~{Ahmed}, R.~{Ferrus}, R.~{Fedrizzi} \emph{et~al.}, ``Towards
  {SDN/NFV-enabled} satellite ground segment systems: Bandwidth on demand use
  case,'' in \emph{Proc. IEEE Int Conf. Commun. Workshops}, Paris, France, Jun.
  2017, pp. 894--899.

\bibitem{7490612}
F.~{Riffel} and R.~{Gould}, ``Satellite ground station virtualization: Secure
  sharing of ground stations using software defined networking,'' in
  \emph{Proc. Annu. IEEE Syst. Conf.}, Orlando, USA, Apr. 2016, pp. 1--8.

\bibitem{7883913}
D.~{Zhou}, M.~{Sheng}, X.~{Wang} \emph{et~al.}, ``Mission aware contact plan
  design in resource-limited small satellite networks,'' \emph{IEEE Trans.
  Commun.}, vol.~65, no.~6, pp. 2451--2466, 2017.

\bibitem{7976296}
M.~{Tipaldi} and L.~{Glielmo}, ``A survey on model-based mission planning and
  execution for autonomous spacecraft,'' \emph{IEEE Syst. J.}, vol.~12, no.~4,
  pp. 3893--3905, 2018.

\bibitem{7272824}
F.~{Perea}, R.~{Vazquez}, and J.~{Galan-Viogue}, ``Swath-acquisition planning
  in multiple-satellite missions: an exact and heuristic approach,'' \emph{IEEE
  Trans. Aerosp. Electron. Syst.}, vol.~51, no.~3, pp. 1717--1725, 2015.

\bibitem{5332322}
E.~{Maurer}, F.~{Mrowka}, A.~{Braun} \emph{et~al.}, ``Terra{SAR}-{X} mission
  planning system: Automated command generation for spacecraft operations,''
  \emph{IEEE Trans. Geosci. Remote Sens.}, vol.~48, no.~2, pp. 642--648, 2010.

\bibitem{6838778}
F.~{Xhafa}, X.~{Herrero}, A.~{Barolli} \emph{et~al.}, ``A tabu search algorithm
  for ground station scheduling problem,'' in \emph{Proc. IEEE Int. Conf. Adv.
  Informa. Netw. Appl.}, Victoria, Canada, May 2014, pp. 1033--1040.

\bibitem{6818281}
Z.~{Xu}, B.~{Lou}, and C.~{Wang}, ``Task scheduling of satellite ground station
  systems based on the neighbor-area search algorithm,'' in \emph{Proc. Int.
  Conf. Nat. Comput.}, Shenyang, China, Jul. 2013, pp. 1830--1834.

\bibitem{6581256}
A.~{Tepe} and G.~{Yilmaz}, ``A survey on cloud computing technology and its
  application to satellite ground systems,'' in \emph{Proc. Int. Conf. Recent
  Adv. Space Technol.}, Istanbul, Turkey, Jun. 2013, pp. 477--481.

\bibitem{7463884}
C.~{Fan}, X.~{Zhao}, L.~{Xie} \emph{et~al.}, ``A resource mapping method in
  cloud-based satellite ground system,'' in \emph{Proc. IEEE Int. Conf.
  SmartCity/SocialCom/SustainCom}, Chengdu, China, Dec. 2015, pp. 1163--1166.

\bibitem{liu2015improve}
J.~Liu, Y.~Li, Y.~Zhang \emph{et~al.}, ``Improve service chaining performance
  with optimized middlebox placement,'' \emph{IEEE Trans. Serv. Comput.},
  vol.~10, no.~4, pp. 560--573, 2015.

\bibitem{IBM-CPLEX-WEB}
\BIBentryALTinterwordspacing
{IBM ILOG CPLEX} optimization studio v12.10.0. [Online]. Available:
  \url{https://www.ibm.com/support/knowledgecenter/SSSA5P_12.10.0/COS_KC_home.html}
\BIBentrySTDinterwordspacing

\bibitem{leiserson1985fat}
C.~E. Leiserson, ``Fat-trees: universal networks for hardware-efficient
  supercomputing,'' \emph{IEEE Trans. Comput.}, vol. 100, no.~10, pp. 892--901,
  1985.

\bibitem{guo2009bcube}
C.~Guo, G.~Lu, D.~Li \emph{et~al.}, ``Bcube: a high performance, server-centric
  network architecture for modular data centers,'' in \emph{Proc. ACM SIGCOMM},
  Barcelona, Spain, Aug. 2009, pp. 63--74.

\bibitem{greenberg2009vl2}
A.~Greenberg, J.~R. Hamilton, N.~Jain \emph{et~al.}, ``{VL2}: a scalable and
  flexible data center network,'' in \emph{Proc. ACM SIGCOMM}, Barcelona,
  Spain, Aug. 2009, pp. 51--62.

\bibitem{raayatpanah2018virtual}
M.~A. Raayatpanah and T.~Weise, ``Virtual network function placement for
  service function chaining with minimum energy consumption,'' in \emph{Pro.
  IEEE Int. Conf. Comput. Commun. Eng. Technol.}, Beijing, China, Aug. 2018,
  pp. 198--202.

\bibitem{tan2017nsga}
B.~Tan, H.~Ma, and Y.~Mei, ``A {NSGA-II}-based approach for service resource
  allocation in cloud,'' in \emph{Proc. IEEE Congr. Evol. Comput.}, San
  Sebastian, Spain, Jun. 2017, pp. 2574--2581.

\bibitem{sun2016forecast}
Q.~Sun, P.~Lu, W.~Lu \emph{et~al.}, ``Forecast-assisted {NFV} service chain
  deployment based on affiliation-aware v{NF} placement,'' in \emph{Proc.
  GLOBECOM}, Washington, USA, Dec. 2016, pp. 1--6.

\bibitem{tang2018dynamic}
H.~Tang, D.~Zhou, and D.~Chen, ``Dynamic network function instance scaling
  based on traffic forecasting and {VNF} placement in operator data centers,''
  \emph{IEEE Trans. Parallel Distrib. Syst.}, vol.~30, no.~3, pp. 530--543,
  2018.

\bibitem{li2018deep}
B.~Li, W.~Lu, S.~Liu, and Z.~Zhu, ``Deep-learning-assisted network
  orchestration for on-demand and cost-effective v{NF} service chaining in
  inter-{DC} elastic optical networks,'' \emph{IEEE/OSA J. Opt. Commun. Netw.},
  vol.~10, no.~10, pp. 29--41, 2018.

\bibitem{jou2018architecture}
B.~T. Jou, O.~Vidal, J.~Cahill \emph{et~al.}, ``Architecture options for
  satellite integration into {5G} networks,'' in \emph{Proc. Eur. Conf. Netw.
  Commun.}, Ljubljana, Slovenia, Jun. 2018, pp. 398--399.

\bibitem{ferrus2016virtualization}
R.~Ferrus, H.~Koumaras, O.~Sallent \emph{et~al.}, ``On the virtualization and
  dynamic orchestration of satellite communication services,'' in \emph{Proc.
  IEEE Veh. Technol. Conf.}, Montreal, Canada, Sep. 2016, pp. 1--5.

\bibitem{ahmed2017satellite}
T.~Ahmed, R.~Ferrus, R.~Fedrizzi \emph{et~al.}, ``Satellite gateway diversity
  in {SDN/NFV}-enabled satellite ground segment systems,'' in \emph{Proc. IEEE
  Int Conf. Commun. Workshops}, Paris, France, May 2017, pp. 882--887.

\bibitem{9051712}
Y.~{Liu}, Y.~{Chen}, Y.~{Jiao} \emph{et~al.}, ``A shared satellite ground
  station using user-oriented virtualization technology,'' \emph{IEEE Access},
  vol.~8, pp. 63\,923--63\,934, 2020.

\bibitem{946526}
J.~E. {Hilland}, R.~R. {Wessen}, D.~{Porter} \emph{et~al.}, ``A market-based
  conflict resolution approach for satellite mission planning,'' \emph{IEEE
  Trans. Eng. Manag.}, vol.~48, no.~3, pp. 272--282, 2001.

\bibitem{8418365}
L.~{Li}, H.~{Chen}, J.~{Li} \emph{et~al.}, ``Preference-based evolutionary
  many-objective optimization for agile satellite mission planning,''
  \emph{IEEE Access}, vol.~6, pp. 40\,963--40\,978, 2018.

\bibitem{5980932}
C.~{Wang}, X.~{Song}, Z.~{Deng} \emph{et~al.}, ``Online system for satellite
  observation planning,'' in \emph{Proc. Int. Conf. Geoinform.}, Shanghai,
  China, Jun 2011, pp. 1--4.

\bibitem{8865283}
P.~{Han}, Z.~{He}, Y.~{Geng} \emph{et~al.}, ``Mission planning for agile earth
  observing satellite based on genetic algorithm,'' in \emph{Proc. Chin.
  Control Conf.}, Guangzhou, China, Jul 2019, pp. 2118--2123.

\bibitem{ahuja2004multi}
R.~K. Ahuja, J.~B. Orlin, S.~Pallottino \emph{et~al.}, ``A multi-exchange
  heuristic for the single-source capacitated facility location problem,''
  \emph{Manage. Sci.}, vol.~50, no.~6, pp. 749--760, 2004.

\bibitem{2017arXiv170200369R}
W.~Rankothge, F.~Le, A.~Russo \emph{et~al.}, ``Data modelling for the
  evaluation of virtualized network functions resource allocation algorithms,''
  \emph{ArXiv}, vol. abs/1702.00369, feb 2017.

\end{thebibliography}
%
%
%

%



\begin{IEEEbiography}[{\includegraphics[width=1in,height=1.25in,clip,keepaspectratio]{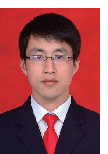}}]{Xiangqiang Gao} received the B.Sc. degree in school of electronic engineering from Xidian University and the M.Sc. degree from Xi\textquoteright an Microelectrinics Technology Institute, Xi\textquoteright an, China, in 2012 and 2015, respectively. He is currently pursuing the Ph.D. degree with the School of Electronic and Information Engineering, Beihang University, Beijing, China. His research interests include rateless codes, software defined network and network function virtualization.\par
\end{IEEEbiography}
\vfill
\begin{IEEEbiography}[{\includegraphics[width=1in,height=1.25in,clip,keepaspectratio]{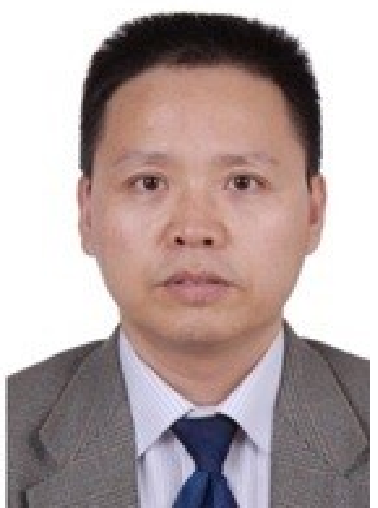}}]{Rongke Liu} received the B.S. and Ph.D. degrees from Beihang University in 1996 and 2002, respectively. He was a Visiting Professor with the Florida Institution of Technology, USA, in 2006; The University of Tokyo, Japan, in 2015; and the University of Edinburgh, U.K., in 2018, respectively. He is currently a Full Professor with the School of Electronic and Information Engineering, Beihang University. He received the support of the New Century Excellent Talents Program from the Minister of Education, China. He has attended many special programs, such as China Terrestrial Digital Broadcast Standard. He has published over 100 papers in international conferences and journals. He has been granted over 20 patents. His research interest covers wireless  communication and space information network.\par
\end{IEEEbiography}
\begin{IEEEbiography}[{\includegraphics[width=1in,height=1.25in,clip,keepaspectratio]{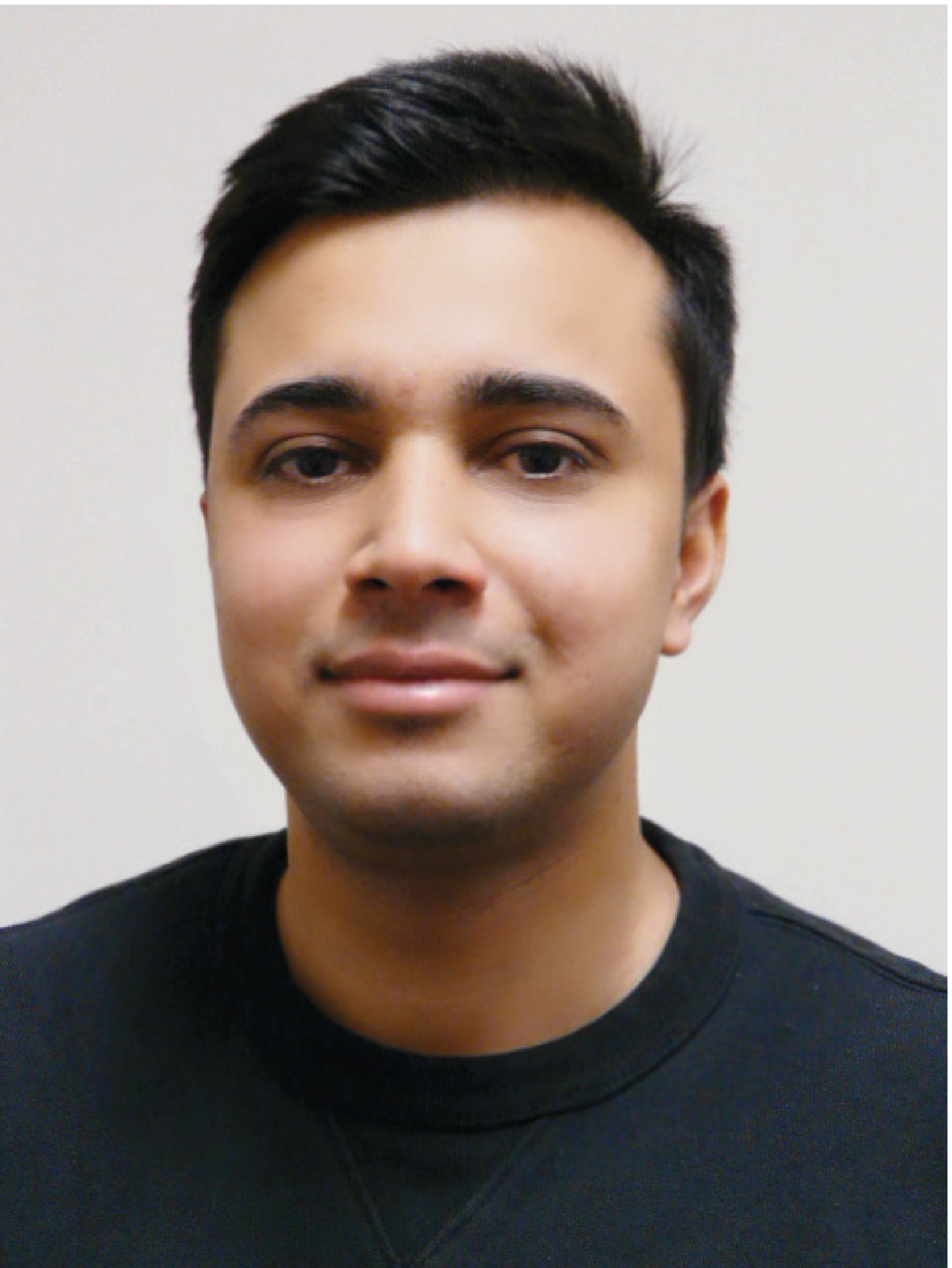}}]{Aryan Kaushik} is currently a Research Fellow in Communications and Radar Transmission at the Institute of Communications and Connected Systems, University College London, United Kingdom. He received PhD in Communications Engineering at the Institute for Digital Communications, School of Engineering, The University of Edinburgh, United Kingdom, in 2020. He received MSc in Telecommunications from The Hong Kong University of Science and Technology, Hong Kong, in 2015. He has held visiting research appointments at the Wireless Communications and Signal Processing Lab, Imperial College London, UK, from 2019-20, the Interdisciplinary Centre for Security, Reliability and Trust, University of Luxembourg, Luxembourg, in 2018, and the School of Electronic and Information Engineering, Beihang University, China, from 2017-19. His research interests are broadly in signal processing, radar, wireless communications, millimeter wave and multi-antenna communications.\par
\end{IEEEbiography}

\vfill


\end{document}